\documentclass[superscriptaddress,amsmath, nofootinbib]{revtex4}   
\usepackage{amsfonts}
\usepackage{euscript}
\usepackage{graphicx}
\usepackage[latin1]{inputenc}
\usepackage{amsmath}
\usepackage{amssymb}
\usepackage[latin1]{inputenc}
\usepackage{amsmath}
\usepackage{amssymb}
\usepackage{amsfonts}
\usepackage{graphicx}
\usepackage[latin1]{inputenc}
\usepackage{amsmath}
\usepackage{amssymb}

\begin{document}


\title{Massless polarized particle and Faraday rotation of light in the Schwarzschild spacetime}
\author{Alexei A. Deriglazov }
\email{alexei.deriglazov@ufjf.edu.br} \affiliation{Depto. de Matem\'atica, ICE, Universidade Federal de Juiz de Fora,
MG, Brazil,} \affiliation{Department of Physics, Tomsk State University, Lenin Prospekt 36, 634050, Tomsk, Russia}


\date{\today}

\begin{abstract}
We present the manifestly covariant Lagrangian of a massless polarized particle, that implies all dynamic and algebraic equations as the conditions of extreme of this variational problem. The model allows for minimal interaction with a gravitational field, leading to the equations, coinciding with Maxwell equations in the geometrical optics approximation. The model allows also a wide class of non minimal interactions, which suggests an alternative way to study the  electromagnetic radiation beyond the leading order of geometrical optics.  As a specific example, we construct a curvature-dependent interaction in Schwarzschild spacetime, predicting the Faraday rotation of polarization plane, linearly dependent on the wave frequency.  As a result, the  Schwarzschild spacetime generates a kind of angular rainbow of light: waves of different frequencies, initially linearly-polarized in one direction, acquire different orientations of their polarization planes when propagate along the same ray.
\end{abstract}

\maketitle 

\section{Introduction.}

In general relativity the electromagnetic radiation, taken in the approximation of a massless point particle, propagates along null geodesics. However, in modern applications \cite{Yu_Peng_Zhang_2020_1, Yu_Peng_Zhang_2019_2, Hamada_2018, Biswas_2020_1, Toshmatov_2020, Nucamendi_2020, Azreg-Ainou_2020, Mukherjee_2020, Bin_Chen_2020, Yunlong_Liu_2020, Sheoran_2020, Kaye_Jiale_Li_2020, Santana_2021, Brax_2020} an actual becomes the task to take into account the polarization degrees of freedom of a light beam. This implies analysis of Maxwell equations in curved spacetime, the task with quite long history, see \cite{Skrotskii_1957, Balazs_1958, Plebanski_1959, Mashhoon_1973} and the recent review \cite{Oancea_2020_1}.   Account of the polarization implies the analysis of effects of two types.   
First, interaction of polarization with gravity could produce the polarization-dependent contributions into the equation of geodesic line, causing deviation of trajectories from the null geodesics. In particular, corrections to the trajectory due to helicity of a circularly-polarized light (photon's spin) are under intensive discussion \cite{Mashhoon_1975, Frolov_2011, Yamamoto_2018,  Dolan_2018, Harte_2019, Armen_2020, Andrey_Shoom_2020, Oancea_2020, Frolov_2020}, and are considered as a gravitational analogue of Magnus or spin Hall effects of light observed in medium \cite{Zeldovich_1992, Zeldovich_1992_2, Bliokh_2004, Bliokh_2015}. Second, interaction affects the evolution of the polarization vector, leading to a number of interesting consequences.  An example is a rotation of polarization plane of a linearly-polarized beam around the direction of its propagation in gravitational fields with rotation \cite{Skrotskii_1957, Balazs_1958, Plebanski_1959, Ishihara_1988, Fayos_1982, Nouri-Zonoz_1999, Yihan_Chen_2011, Connors_1980, Sereno_2004, Ghosh_2016, Schneiter_2018}. This is considered as a gravitational analogue of the Faraday effect \cite{Rybicki_1979, Landau_8}. 
  
Various approaches have been used to theoretically describe and analyze the polarization degrees of freedom and related effects.  Among them are the one-particle quantum-mechanical interpretation of  Maxwell equations \cite{Akhiezer_1965, Bialynicki_1996, Bialynicki_2019} and of QFT \cite{Berard_2004, Gitman_2015, Mohrbach_2006, Kosinski_2016, Kosinski_2020_1, Kosinski_2020_2, Dodin_2015_1, Dodin_2015_2, Mohrbach_2007, Duval_2006}, as well as the geometrical optics approximation to Maxwell equations in curved space-time \cite{Born_2005, Skrotskii_1957, Balazs_1958, Plebanski_1959, Frolov_2011, Yamamoto_2018, Dolan_2018, Harte_2019, Armen_2020, Andrey_Shoom_2020, Oancea_2020, Frolov_2020, Ishihara_1988, Fayos_1982, Nouri-Zonoz_1999, Yihan_Chen_2011, Connors_1980, Sereno_2004, Ghosh_2016, Schneiter_2018}. 
In the geometrical optics, solutions to the wave equations are sought in the form  $A=\mbox{Re}(a e^{iS})$. As $A$ can be taken either the electro-magnetic field $F_{\mu\nu}$ \cite{Dolan_2018, Frolov_2011, Mieling}, or its vector potential $A^\mu$ \cite{Misner_2011}. The resulting partial differential equations can be split in two parts: the (eikonal) equation for the phase $S$, $H(x, \nabla S)=0$, and the equations of the form $F(x, \nabla a, \nabla S, \ldots )=0$. The latters can be used to determine the polarization vector $a$ when the phase is already known.
The solution of eikonal equation can be reduced to the solution of a Hamiltonian equations of classical 
mechanics \cite{Maslov_1988, Arnold_1989}. Thereby, with an electromagnetic wave can be associated the congruence of world lines that are solutions to these Hamiltonian equations. In the geometrical optics approximation, they turn out to be  null geodesics, while the polarization vector undergoes parallel transport  along these lines \cite{Misner_2011, Frolov_2011}. Maxwell equations in an arbitrary curved background were analyzed in the geometrical optics and weak field approximations in \cite{Skrotskii_1957, Balazs_1958, Plebanski_1959}. It was observed that angular velocity of the polarization plane is 
due to $g_{0i}$\,-components  of the metric. Hence the Faraday rotation is not expected in the Schwarzschild space, but may occur in Kerr and other rotational spaces, if the precession is an accumulative effect during the wave propagation. Computing the total rotation angle in the leading approximation, Plebansky found \cite{Plebanski_1959} that polarization remains unchanged for the wave that does not  penetrate a rotating matter. However, taking into account the higher-order corrections, the  Faraday rotation has been predicted in Kerr and other spaces \cite{Ishihara_1988, Fayos_1982, Nouri-Zonoz_1999, Yihan_Chen_2011, Connors_1980}. Importantly, for the gravitational Faraday rotation discussed in the literature, the rotation angle does not depend on the wave frequency. 

The eikonal equation in geometrical optics turns out to be closed equation for determining the phase $S$. As a consequence, the corresponding Hamiltonian equations for light rays do not involve the polarization vector. In particular,  the photon's  trajectory does not depend on its helicity. 
In recent works \cite{ Frolov_2011, Oancea_2020, Andrey_Shoom_2020, Frolov_2020}  were proposed the modifications, allowing to include the polarization vector into the eikonal equation. This implies equations for the trajectory with a dependence on helicity, that is, predicting the gravitational spin-Hall effect.

In the above mentioned approaches, with the light beam we associate a kind of massless particle, endowed with extra degrees of freedom describing its  polarization or helicity. Therefore, an interesting task is to develop, in a systematic form, the manifestly covariant description of a massless polarized particle in curved spacetime. In the case of a massive particle with spin, this is achieved either with help of the Mathisson-Papapetrou-Tulczyjew-Dixon (MPTD) equations of a rotating body in general relativity \cite{Dixon_1964}, or in the framework of vector model of spinning particle, see \cite{AAD_Rec, AAD_2019} and references therein. It would be natural to take the massless limit of these theories. However, this turns out to be problematic. First, MPTD equations in the massless limit contain denominators proportional to space-time curvature \cite{Duval_2017}, and the flat-space limit becomes problematic\footnote{For instance, in Schwarzschild metric the term $\frac{S^{\mu\rho}R_{\rho\nu\alpha\beta}S^{\alpha\beta}P^\nu}{S^{\mu\nu}R_{\mu\nu\alpha\beta}S^{\alpha\beta}}$ does not depend on the Schwarzschild mass.}.  Second, 
MPTD equations imply wrong dependence of acceleration on velocity \cite{AAD_2015, AAD_2019}, that diverges when $v\rightarrow c$, so they hardly can be expected to describe the ultra relativistic particles or photons. The vector model with null gravimagnetic moment is equivalent to MPTD equations \cite{AAD_Rec}, and hence suffers from the same problems.  The wrong behavior of MPTD equations was improved by adding a non minimal interaction through unit gravimagnetic moment \cite{AAD_2015, AAD_2016}. The gravimagnetic spinning particle has a reasonable ultra-relativistic behavior, but its massless limit still remains problematic. A key role in this model play a pair of second class constraints, with the Poisson bracket $\{T_1, T_2\}\sim p^2\sim m^2$. In the massless limit we obtain $\{T_1, T_2\}=0$, that is the second-class constraints turn into the first-class, the latter remove two more degrees of freedom from the physical sector. In the result, in the massless limit there are no of physical degrees of freedom in the spin-sector. While this limiting case could be used to describe helicity \cite{Carapulin_2020} (and therefore its influence on the trajectory), it fails to describe the evolution of the polarization degrees of freedom.  Another drawback of both theories is that their connection with Maxwell's equations is not clear. 

As the formalism of massive spinning particles hardly can be adapted to the massless case, we can try to develop the model of a massless polarized particle in an independent way. Below we propose one possible version of such a theory, and see to what extent it is able to capture the properties of light propagation in curved spacetime. 

The work is organized as follows. In Sect. \ref{ss2}, with a plane monochromatic wave we associate a massless polarized particle, and discuss the variables and constraints appropriate for its description. This allows us to give an exact formulation of the problem under discussion, see the end of the section. It is known that the inclusion of interaction in a free theory with Dirac constraints generally represents a non trivial task. 
Nevertheless, in Sect. \ref{ss3} we present two equivalent Lagrangian actions of the massless polarized particle, which allow for the minimal interaction with an arbitrary gravitational field. We give their Hamiltonian formulation and discuss the physical sector of the theory.  In Sect. \ref{ss4} we show that the model admits a wide class of non minimal polarization-curvature interactions, and show that parallel transport of the polarization tensor is disturbed by the  curvature-dependent terms. 
One specific example of the non minimal interaction is discussed in some details in Sect. \ref{ss5}. The obtained  equations predict linearly dependent on the wave frequency Faraday rotation of the polarization plane even in the Schwarzschild spacetime.  In the concluding section we discuss a number of unusual properties of our model, considered as a Hamiltonian system with Dirac constraints. Appendix 1 contains some technical details and was included to make the work a self-contained. In Appendix 2 we present detailed analysis of the photon's Lagrangian with three auxiliary variables.

{\bf Notation.}  We use Greek letters for indexes of four-dimensional quantities  and Latin letters for the three-dimensional quantities. Greek and Latin letters from the beginning of the alphabet are used for indexes in Minkowski space, $x^\alpha=(x^0, x^a)$, with the metric $\eta_{\alpha\beta}$ of signature $(-, +, +, +)$.  Greek and Latin letters from the middle of the alphabet are used for indexes in curved space, $x^\mu=(x^0, x^i)$.  Our variables are taken in an arbitrary parameterization $\tau$, and we denote $\dot x^\mu=\frac{dx^\mu}{d\tau}$.
For the four-dimensional quantities we suppress the
contracted indexes and use the notation  $N^\mu{}_\nu\dot
x^\nu=(N\dot x)^\mu$, $\omega^2=g_{\mu\nu}\omega^\mu\omega^\nu$, and so on.  
Suppressing the indexes of three-dimensional vectors, we use bold letters. The Euclidean scalar and vector products are $({\bf E}, {\bf B})=E_aB_a$, $[{\bf E}, {\bf B}]_a=\epsilon_{abc}E_b B_c$.

\section{Basic variables and constraints for describing a massless polarized particle.}\label{ss2}
We start from solutions to the Maxwell equations in empty space in the form of plane monochromatic waves, which can be written as follows (see \cite{Born_2005, Landau_2} or Appendix 1):
\begin{eqnarray}\label{pol.2}
{\bf E}(x^\alpha)= 
{\bf e}_1\cos \frac{\tilde\omega}{c}(\hat{\bf k}{\bf x}-ct) -\epsilon{\bf e}_2\sin \frac{\tilde\omega}{c}(\hat{\bf k}{\bf x}-ct), \qquad 
{\bf B}(x^\alpha)=[\hat{\bf k}, {\bf E}],  
\end{eqnarray}
The unit vector $\hat{\bf k}$ points the direction orthogonal to a wave front, $\tilde\omega>0$ is  frequency of the wave\footnote{We use the notation $\tilde\omega$ to distinguish the frequency from the basic variables $\omega^\mu$ that will appear below.},  then $T=2\pi/\tilde\omega$ gives period, and $\lambda=cT$ is the wavelength. The columns ${\bf e}_1$ and ${\bf e}_2$ are three-dimensional constant vectors. The set $(\hat{\bf k}, {\bf e}_2, {\bf e}_1)$  consists of mutually orthogonal vectors, which form the right-handed triad. The set $(\hat{\bf k}, {\bf E}, {\bf B})$  consists of mutually orthogonal vectors, which also form the right-handed triad, and $|{\bf E}|=|{\bf B}|$.  If we look at the plane of ${\bf E}$ and ${\bf B}$ from the end of the vector $\hat{\bf k}$, the vectors ${\bf E}(t)$ and ${\bf B}(t)$ at given point ${\bf x}$ rotate clockwise when $\epsilon=+1$, and counterclockwise when $\epsilon=-1$. Let us  fix $t$ and consider the straight line through ${\bf x}$ in the direction of $\hat{\bf k}$, that is, consider the instantaneous configuration of the wave along the straight line ${\bf x}+\hat{\bf k}s$, $s\in{\mathbb R}$. Then the ends of the vectors ${\bf E}$ and ${\bf B}$ lie on the surface of the elliptical cylinder with semi-axes $|{\bf e}_1|$ and $|{\bf e}_2|$, and make one revolution after the increment  $\triangle s=cT$. When one of the vectors ${\bf e}_i$ vanishes, ${\bf E}$ oscillates along the another vector, and we have the linearly polarized wave.

Introducing the four-dimensional wave vector 
\begin{eqnarray}\label{pol.2.1}
k^\alpha=(k^0, {\bf k})=(\frac{\tilde\omega}{c}, \frac{\tilde\omega}{c}\hat{\bf k}), \qquad k^2=0, 
\end{eqnarray}
the phase can be written in the Lorentz-invariant form: $\frac{\tilde\omega}{c}(\hat{\bf k}{\bf x}-ct)=\eta_{\alpha\beta}k^\alpha x^\beta$. 

Let's choose some space-time point $x^\alpha_0=(ct_0, {\bf x}_0)$, and denote ${\bf E}(x^\alpha_0)\equiv {\bf E}_0$, ${\bf B}(x^\alpha_0)\equiv {\bf B}_0$. Consider the chain of events 
\begin{eqnarray}\label{pol.4}
x^\alpha(t)=(c(t_0+t), ~ {\bf x}_0+c\hat{{\bf k}}t), \qquad t\in[0, t_1].
\end{eqnarray}
We identify them with world line of a particle  with velocity equal to ${\bf v}=d{\bf x}/dt=c \hat{{\bf k}}$, and the speed equal to the speed of 
light, $|d{\bf x}/dt|=c$.  The electric and magnetic fields of a plane wave (\ref{pol.2}) turn out to be constant along the world-line: ${\bf E}(x^\alpha(t))={\bf E}_0$, ${\bf B}(x^\alpha(t))={\bf B}_0$.  By this, with  given plane wave we associate a particle, that moves along the  null world-line (\ref{pol.4}) and carries
a pair of mutually ortogonal constant vectors ${\bf E}_0$ and ${\bf B}_0$, the latter lie on the plane ortogonal to the velocity vector, and such that  $({\bf v}, {\bf E}_0, {\bf B}_0)$ is the right-handed triad. We call it a massless polarized particle or, in short, a photon. By construction, it describes the evolution of electric and magnetic fields of the plane wave (\ref{pol.2}) along the world-line (\ref{pol.4}).

Note, that with clockwise and counterclockwise waves is associated the same particle, that is helicity of a photon is not taken into account.  In addition, the triad $({\bf v}, {\bf E}_0, {\bf B}_0)$ no longer contains an information about the wave frequency, and its dynamics in Minkowski space is trivial. However, if  our particle can interact with gravity,  this simple model can be useful for discussing some problems of current interest in the field of black hole physics: (a) What type of interaction with gravity does the photon allow? (b) Influence of a polarization-gravity interaction on the trajectory. (c) Dynamics of the polarization plane of the wave propagating in a gravitational background. 

To describe the trajectory of a photon in a manifestly covariant way, we take it in a parametric form: ${\bf x}(t)\rightarrow x^\alpha(\tau)=(ct(\tau), ~ {\bf x}(\tau)~ )$, requiring that directions of time and of evolution parameter coincide (see also Appendix 1)
\begin{eqnarray}\label{best}
\dot x^0> 0.
\end{eqnarray}
Further, the only simple way to describe covariantly a triad of orthogonal three-vectors is to consider the following system of constraints: 
\begin{eqnarray}\label{pol.5}
p^2=0, \qquad f_{\alpha\beta}p^\beta=0, \qquad \tilde f_{\alpha\beta}p^\beta=0, 
\end{eqnarray}
where $\tilde f_{\alpha\beta}=\epsilon_{\alpha\beta\gamma\delta}f^{\gamma\delta}$ with $\epsilon_{0123}=-1$  is the tensor dual to $f_{\alpha\beta}$. Then the three-dimensional quantities 
\begin{eqnarray}\label{pol.6}
E_a\equiv f_{a0}, \qquad B_a\equiv\frac12\epsilon_{abc}f_{bc}, \quad \mbox{then} \quad f_{ab}=\epsilon_{abc}B_c,
\end{eqnarray}
obey the desired properties: ${\bf E}, {\bf B}$ and ${\bf p}$ are mutually orthogonal, ${\bf B}=[\hat{\bf p}, {\bf E}]$, so the triad $(\hat{\bf p}={\bf p}/p^0, {\bf E}, {\bf B})$ is right-handed. Due to the constraints (\ref{pol.5}), the tensor $f$ has only two independent components. 

The constraints (\ref{pol.5}) can be "solved" in terms of  space-like vector $\omega^\alpha$ orthogonal to $p_\alpha$. Indeed, if we require
\begin{eqnarray}\label{pol.7}
p^2=0, \qquad (\omega, p)=0 \qquad \omega^2>0,
\end{eqnarray}
the quantity
\begin{eqnarray}\label{pol.8}
f_{\alpha\beta}=p_\alpha\omega_\beta-p_\beta\omega_\alpha,
\end{eqnarray} 
obeys to the conditions (\ref{pol.5}). The fields then can be presented through $p$ and ${\boldsymbol{\omega}}$ as follows
\begin{eqnarray}\label{pol.9}
{\bf E}=p^0 {\boldsymbol{\omega}}_{tr}, \qquad {\bf B}=[{\bf p}, {\boldsymbol{\omega}}]\equiv [{\bf p}, {\boldsymbol{\omega}}_{tr}],
\end{eqnarray} 
where ${\boldsymbol{\omega}}_{tr}=N({\bf p}){\boldsymbol{\omega}}$ is the transverse part of ${\boldsymbol{\omega}}$, that is the projection of ${\boldsymbol{\omega}}$ on the plane ortogonal to ${\bf p}$ with help of the projector $N({\bf p})_{ab}=\delta_{ab}-\frac{p_a p_b}{{\bf p}^2}$. The tensor $f$ is invariant under the gauge transformation $\omega_\alpha\rightarrow \omega_\alpha+\alpha(\tau)p_\alpha$. We will construct our theory in terms of  variables $x^\alpha$ and $\omega^\alpha$  instead of $x^\alpha$ and $f_{\alpha\beta}$, that turns out to be an easier task. Denote $p_\alpha$ and $\pi_\alpha$ the conjugate momenta for $x^\alpha$ and $\omega^\alpha$.  It is clear from Eqs. (\ref{pol.7}) and (\ref{pol.8})  that we need a theory with rather unusual properties, where the variable $\omega$ is involved in the construction of the physical sector, but its momentum is not. Models with such properties can be easily constructed  using the Grassmann variables \cite{Berezin_1977}. Without them, as we will see, this represents a rather nontrivial task.

Taking all the above into account, we can now formulate in a more concrete form the problem of constructing a model of massless polarized particle.
We need a variational problem with the Lagrangian $L(x^\alpha, \omega^\alpha)$, that has the following properties. \par 
\noindent (1) All expected constraints (in particular, (\ref{pol.7})), together with dynamic equations, should appear as the extreme conditions of the 
variational problem. \par 
\noindent (2) We need such a set of constraints, that in Hamiltonian formalism the physical sector contains the observables ${\bf x}(t)$,  ${\bf p}(t)$ and $f_{\alpha\beta}(t)$. According to Dirac \cite{Dirac_1977}, by observables we mean the variables with unambiguous dynamics, see also the discussion below Eq.(\ref{lag.22}).  \par 
\noindent (3) The theory should allow interaction with gravity.

\section{Lagrangian of massless polarized particle in curved space-time.}\label{lagrangian}\label{ss3}
Massless point particle is described starting from the variational problem with one auxiliary variable, see Appendix 1. To take into account polarization, we need more auxiliary variables.  Here I propose two different Lagrangian actions, the first with three and another with four auxiliary variables. They have the same physical sector, so any one of them can be used to describe the  massless polarized particle in curved spacetime.  I was not able to find a formulation with less then three auxiliary variables. 

In the first case, our dynamical variables of configuration space with Minkowski metric $\eta_{\alpha\beta}=(-, +, +, +)$ are: $x^\alpha(\tau), \omega^\alpha(\tau), e_1(\tau), e_2(\tau), e_3(\tau)$, $\alpha=0, 1, 2, 3$. Under the Poincar\'e transformations, $x$ behave as the position variable: $x'^\alpha=\Lambda^\alpha{}_\beta x^\beta+a^\alpha$, and $\omega$ is the vector-like variable: $\omega'^\alpha=\Lambda^\alpha{}_\beta\omega^\beta$. It can be considered a mechanical analogue of the vector potential $A^\mu(x^\nu)$ of electromagnetic field.  The auxiliary variables $e_i$ are scalar functions. Dynamics of the theory is determined by variational problem with the following Lagrangian action: 
\begin{eqnarray}\label{lag.1}
S=\int d\tau\frac{1}{4e_1}\left[\dot x^2+\frac{1-e_3}{1+e_3}\frac{(\dot x\omega)^2}{\omega^2}+e_2\dot\omega^2-\sqrt{\left(\dot xN\dot x+e_2\dot\omega
N\dot\omega\right)^2-4e_2(\dot xN\dot\omega)^2}\right], 
\end{eqnarray}
where the matrix $N$ is projector on the hyperplane orthogonal to $\omega$
\begin{equation}\label{lag.2}
N_{\alpha\beta}= \eta_{\alpha\beta}-\frac{\omega_\alpha \omega_\beta}{\omega^2}, \quad \mbox{then} \quad N^2=N, \quad N_{\alpha\beta} \omega^\beta=0.
\end{equation} 
As $\omega^2$ enters into the denominator, we assumed $\omega^2\ne 0$. We also assume $\dot x^2\leq 0$, then in the (spinless) limit, $e_2\to 0$, $e_3\to 1$,
the functional (\ref{lag.1}) reduces to the standard Lagrangian action 
of a massless point particle, $\int d\tau\frac{1}{2e}\dot x^2$. It can also be compared with that of massive spinning particle \cite{AAD_Rec}
\begin{eqnarray}\label{lag.3}
S=\int d\tau\frac{1}{4e}\left[\dot xN\dot x+\dot\omega N\dot\omega-\sqrt{\left[\dot xN\dot x+\dot\omega
N\dot\omega\right]^2-4(\dot xN\dot\omega)^2}\right]- \frac{e}{2}(m^2c^2-\frac{\alpha}{\omega^2}).
\end{eqnarray}
The action (\ref{lag.1}) does not represent a massless limit of (\ref{lag.3}).  To obtain it  from (\ref{lag.3}), we need to replace the first term $\dot xN\dot x=\dot x^2-(\dot x \omega)^2/\omega^2$ on $\dot x^2+(1-e_3)(\dot x \omega)^2/(1+e_3)\omega^2$, to remove the projector $N$ from the second term; to insert in various places the auxiliary variable $e_2$; and then to take the limit $m\to 0$ and $\alpha\to 0$. In the result, the two theories turn out to be rather different: in the model of massive spinning particle there are three constraints of first class and two of second class, while the massless polarized particle is subject to four constraints of first class. 

To formulate the theory (\ref{lag.1}) in a curved space-time with the metric $g_{\mu\nu}(x^\rho)$, we fix the position of curve indexes as follows: $x^\mu$ and $\omega^\mu$, and then introduce the minimal interaction, that is we replace 
$\eta_{\alpha\beta}\rightarrow g_{\mu\nu}$, and usual derivative of $\omega$ by the covariant one,
\begin{equation}\label{lag.4}
\dot\omega^\alpha\rightarrow \nabla\omega^\mu=\dot\omega^\mu +\Gamma^\mu{}_{\rho\sigma}\dot
x^\rho\omega^\sigma.
\end{equation}
This gives the action 
\begin{eqnarray}\label{lag.5}
S=\int d\tau\frac{1}{4e_1}\left[\dot x^2+\frac{1-e_3}{1+e_3}\frac{(\dot x\omega)^2}{\omega^2}+e_2(\nabla\omega)^2-\sqrt{\left(\dot xN\dot x+e_2\nabla\omega
N\nabla\omega\right)^2-4e_2(\dot xN\nabla\omega)^2}\right], 
\end{eqnarray}
where $\dot x^2$ means now $g_{\mu\nu}\dot x^\mu\dot x^\nu$, $\omega_\mu=g_{\mu\nu}\omega^\nu$, and so on. 
Velocities $\dot x^\mu$, $\nabla\omega^\mu$ and projector $N_{\mu\nu}$ transform like contravariant vectors and
covariant tensor, so the action is manifestly invariant under the general-coordinate transformations.

In the Appendix 2 we show that the functional (\ref{lag.5}) implies all the desired constraints (\ref{pol.5}) which, together with dynamical equations, arise as the conditions of extreme of this variational problem. Hence the theory describes a photon in an arbitrary gravitational background. Here we discuss a more simple action, with four auxiliary variables $\tilde e_1, \tilde e_2, \tilde e_3,$ and  $\tilde e_4$
\begin{eqnarray}\label{lag.6}
S=\int d\tau ~ \frac{1}{2\tilde e_1}(\dot x^\alpha+\tilde e_4\dot\omega^\alpha)\left[\eta_{\alpha\beta}+\tilde e_3\frac{\omega_\alpha\omega_\beta}{\omega^2}\right](\dot x^\beta+\tilde e_4\dot\omega^\beta)+\frac{1}{2\tilde e_2}\dot\omega^2. 
\end{eqnarray}
Invariance of the action under Poincare transformations gives two Noether charges, they are conjugated momentum for $x^\mu$, and total angular momentum: $p_\alpha$, $ J_{\alpha\beta}=x_{[\alpha}p_{\beta]}+ \omega_{[\alpha}\pi_{\beta]}$. Here $p_\alpha=\frac{\partial L}{\partial\dot x_\alpha}$ and $\pi_\alpha=\frac{\partial L}{\partial\dot\omega^\alpha}$. Note that $x_{[\alpha}p_{\beta]}$ and $\omega_{[\alpha}\pi_{\beta]}$  are not preserved separately. Note also that $p^\alpha$ and $\pi^\alpha$ are not proportional to $\dot x^\alpha$ and $\dot\omega^\alpha$, see Eqs. (\ref{lag26}) below. 
Introducing the minimal coupling in (\ref{lag.6}), we obtain the action
\begin{eqnarray}\label{lag.7}
S=\int d\tau ~ \frac{1}{2\tilde e_1}(\dot x^\mu+\tilde e_4\nabla\omega^\mu)\left[g_{\mu\nu}+\tilde e_3\frac{\omega_\mu\omega_\nu}{\omega^2}\right](\dot x^\nu+\tilde e_4\nabla\omega^\nu)+\frac{1}{2\tilde e_2}(\nabla\omega)^2. 
\end{eqnarray}

{\bf Hamiltonian formulation.} To study the theory (\ref{lag.7}), we use the Hamiltonian formalism, which is well adapted for the analysis of  a constrained theory \cite{Dirac_1977, Gitman_1990, AAD_Book}. 
Conjugate momenta for $x^\mu$, $\omega^\mu$ and $\tilde e_i$  are
denoted as $p_\mu$, $\pi_\mu$ and $p_{ei}$.  Since
$p_{ei}=\frac{\partial L}{\partial\dot{\tilde e}_i}=0$, the momenta $p_{ei}$ represent the trivial primary constraints,
$p_{ei}=0$. Further, $\pi_\mu=\frac{\partial L}{\partial\dot\omega^\mu}=\frac{\partial L}{\partial D\omega^\nu}=\frac{1}{\tilde e_2}\nabla\omega_\mu+ \frac{\tilde e_4}{\tilde e_1}\left[g_{\mu\nu}+\tilde e_3\frac{\omega_\mu\omega_\nu}{\omega^2}\right](\dot x^\nu+\tilde e_4\nabla\omega^\nu)$. Computing $p_\mu$, we need to take into account that $\dot x$ enters into $\nabla\omega$, so $p_\mu=\frac{\partial L}{\partial\dot x^\mu}+\frac{\partial L}{\partial D\omega^\nu}\frac{\partial D\omega^\nu}{\partial \dot x^\mu}=\frac{1}{e_1}\left[g_{\mu\nu}+\tilde e_3\frac{\omega_\mu\omega_\nu}{\omega^2}\right](\dot x^\nu+\tilde e_4\nabla\omega^\nu)+\pi_\nu\Gamma^\nu{}_{\mu\rho}\omega^\rho$. We see that $p_\mu$ is not a covariant object, so it is convenient to introduce the canonical momentum
\begin{eqnarray}\label{lag.8}
{\cal P}_\mu= p_\mu-\pi_\sigma\Gamma^\sigma{}_{\mu\rho}\omega^\rho.
\end{eqnarray}
Contrary to $p_\mu$, the canonical momentum is a four vector, so we expect that Hamiltonian and equations of motion will be written in terms of this quantity. The obtained expressions for the momenta imply the relations
\begin{eqnarray}\label{lag.9}
{\cal P}_\mu=\frac{1}{\tilde e_1}\left[g_{\mu\nu}+\tilde e_3\frac{\omega_\mu\omega_\nu}{\omega^2}\right](\dot x^\nu+\tilde e_4\nabla\omega^\nu), \qquad 
\pi_\mu=\frac{1}{\tilde e_2}\nabla\omega^\mu+\tilde e_4{\cal P}_\mu,
\end{eqnarray}
which can be solved with respect to velocities as follows:
\begin{eqnarray}\label{lag.10}
\dot x^\mu=(\tilde e_1+\tilde e_2\tilde e_4){\cal P}^\mu-\tilde e_2\tilde e_4\pi^\mu+\frac{\tilde e_1\tilde e_3}{1+\tilde e_3}\frac{{\cal P}\omega}{\omega^2}\omega^\mu, 
\qquad \nabla\omega^\mu=\tilde e_2(\pi^\mu-\tilde e_4 {\cal P}^\mu).
\end{eqnarray}
Hamiltonian is obtained excluding velocities from the expression
\begin{eqnarray}\label{lag.11}
H=p\dot x+\pi\dot\omega+p_{ei}\dot{\tilde e}_i-L+\lambda_ip_{ei}\equiv{\cal P}\dot x+\pi \nabla\omega+p_{ei}\dot{\tilde e}_i-L+\lambda_ip_{ei},
\end{eqnarray}
where $\lambda_i$ are the Lagrangian multipliers for the primary constraints.  Substituting  (\ref{lag.10})  into (\ref{lag.11}), we obtain the Hamiltonian. Then the Hamiltonian form of the variational problem (\ref{lag.7}) reads
\begin{eqnarray}\label{lag.12}
S_H=\int d\tau ~ p\dot x+\pi\dot\omega+p_{ei}\dot{\tilde e}_i-H= \qquad \qquad \qquad \qquad \qquad \qquad \cr \int d\tau ~  p\dot x+\pi\dot\omega+p_{ei}\dot{\tilde e}_i-
\left[\frac12 (\tilde e_1+\tilde e_2\tilde e_4){\cal P}^2+\frac{\tilde e_2}{2}\pi^2+\tilde e_2\tilde e_4{\cal P}\pi+\frac{\tilde e_1\tilde e_3}{2(1+\tilde e_3)\omega^2}({\cal P}\omega)^2+\lambda_ip_{ei}\right]. 
\end{eqnarray}
The fundamental Poisson brackets $\{x^\mu, p_\nu\}=\delta^\mu{}_\nu$ and $\{\omega^\mu, \pi_\nu\}=\delta^\mu{}_\nu$ imply rather  complicated brackets for ${\cal P}$
\begin{eqnarray}\label{lag.13}
\{x^\mu, {\cal P}_\nu\}=\delta^\mu{}_\nu, \quad \{{\cal P}_\mu, {\cal P}_\nu\}=R_{\mu\nu\rho\delta}\pi^\rho\omega^\delta\equiv\Theta_{\mu\nu}(x, \pi, \omega), \quad
\end{eqnarray}
\begin{eqnarray}\label{lag.14}
\{\omega^\mu, {\cal P}_\nu\}=-\Gamma^\mu{}_{\nu\rho}\omega^\rho, \quad \{\pi_\mu, {\cal P}_\nu\}=\pi_\rho\Gamma^\rho{}_{\mu\nu}, \quad  \{g_{\mu\nu},  {\cal P}_\rho\}=\partial_\rho g_{\mu\nu}, 
\end{eqnarray}
where $R_{\mu\nu\rho\delta}$ is the tensor of curvature and $\Gamma^\mu{}_{\nu\rho}$ is the Christoffel connection. The equations governing the evolution can be obtained either by variation of (\ref{lag.12}) with respect to all variables, or according to the Hamilton's prescription: $dq/d\tau=\{ q, H\}$. 

Preservation in time of primary constraints, $\dot p_{ei}=\{p_{ei}, H\}=0$, gives the algebraic equations of second stage of the Dirac procedure. They imply that all solutions of the variational problem (if any) lie on the constraints surface 
\begin{eqnarray}\label{lag.15}
{\cal P}^2=0,  \quad {\cal P}\pi=0, \quad \pi^2=0,  \label{lag.15} \\ {\cal P}\omega=0. \qquad \qquad ~  \label{lag.15.1}
\end{eqnarray}
We deal with a rather exotic variational problem, leading to the Hamiltonian that contains the square of a constraint, $\frac{\tilde e_1\tilde e_3}{2(1+\tilde e_3)\omega^2}({\cal P}\omega)^2$. This term can not contribute to equations of motion, and we can omite it from the Hamiltonian. This turns out to be crucial for obtaining the theory with desired properties. It is not difficult to construct the Lagrangian action which gives the Hamiltonian containing the constraint ${\cal P}\omega$ instead of its square. This gives a different theory, with the physical sector without a reasonable interpretation.  We also rename the auxiliary variables as follows: $\tilde e_1+\tilde e _2\tilde e_4=e_1$, $\tilde e_2=e_2$, $\frac{\tilde e_1\tilde e_3}{1+\tilde e_3}=e_3$, $\tilde e_2\tilde e_4=e_4$. Then the expression  
\begin{eqnarray}\label{lag.16}
H=\frac{e_1}{2}{\cal P}^2+\frac{e_2}{2}\pi^2+e_4{\cal P}\pi, 
\end{eqnarray}
can be equally used as the Hamiltonian of our theory.  Importantly, to describe all the trajectories, it is sufficient to work with the positively defined dynamical variable $e_1(\tau)>0$, then ${\cal P}^0>0$, see Appendix 1.

For the latter use, let us obtain the algebraic equations of third stage of the Dirac procedure. The constraint 
$\pi^2$ is of first class, so it automatically preserved in time. For the remaining constraints we have
\begin{eqnarray}\label{lag.17}
\{{\cal P}\omega, H\}=0 ~\Rightarrow ~ e_1\omega\Theta{\cal P}+e_4\omega\Theta\pi=0, \quad
\{{\cal P}\pi, H\}=0 ~\Rightarrow ~ e_1\pi\Theta{\cal P}=0, \quad
\{{\cal P}^2, H\}=0 ~\Rightarrow ~ e_4\pi\Theta{\cal P}=0, 
\end{eqnarray}
that is we obtained two more constraints. 
They are absent in Minkowski space, and their appearance in interacting theory would mean its inconsistency. 
As we show below, the constraints surface (\ref{lag.15}) consist of four regions. Only in one of them our theory admits a self-consistent interaction with an arbitrary gravitational field. In this region the third-stage constraints  will be  automatically satisfied. 

When evaluating the expressions (\ref{lag.17}), appear many terms containing the connection $\Gamma$ and derivatives of the metric. They all cancel out with each other in such a way that the resulting expression is a scalar function. The following identities: 
\begin{eqnarray}\label{llag.18}
\partial_\rho g_{\mu\nu}\omega^\mu\omega^\nu=2\omega_\mu\Gamma^\mu{}_{\rho\nu}\omega^\nu, \quad \omega^\mu\partial_\mu g_{\rho\nu}\omega^\nu=2\Gamma_{\rho, \mu\nu}\omega^\mu\omega^\nu, \qquad \qquad \cr \pi^\delta(\partial_\mu g_{\delta\nu}-\Gamma_{\delta, \mu\nu}){\cal P}^\nu={\cal P}_\nu\Gamma^\nu{}_{\mu\delta}\pi^\delta, \quad \nabla g_{\mu\nu}=0, \quad \partial_\rho g^{\mu\nu}{\cal P}_\mu\omega_\nu=-\partial_\rho g_{\mu\nu}{\cal P}^\mu\omega^\nu, 
\end{eqnarray}
are useful  for testing these cancellations. The same happens when obtaining the Hamiltonian equations $dq/d\tau=\{ q, H\}$, they are\footnote{There are also the equations for auxiliary variables: $\dot e_i=\lambda_i$, $\dot p_{ei}=0$. They have a simple meaning: dynamics of $e_i$ is not at all fixed, therefore, we do not write them out further.}
\begin{eqnarray}\label{lag.19}
\dot x^\mu=e_1{\cal P}^\mu+e_4\pi^\mu, \label{lag.19.1} \qquad \qquad \qquad \quad \\ \nabla{\cal P}_\mu=\Theta_{\mu\nu}(e_1{\cal P}^\nu+e_4\pi^\nu)\equiv\Theta_{\mu\nu}\dot x^\nu, \label{lag.19.2} \\
\nabla\omega^\mu=e_2\pi^\mu+e_4{\cal P}^\mu, \label{lag.19.3} \qquad \qquad \qquad \\ \nabla\pi_\mu=0. \qquad \qquad  \qquad \qquad  \quad \qquad  \label{lag.19.4}  
\end{eqnarray}
All non-covariant terms either cancel out or covariantize the derivative: $d/d\tau\to\nabla$. Taking this into account, in practical calculations of Poisson brackets we can ignore the brackets (\ref{lag.14}). 

To analyze the obtained  theory, we start from the constraints (\ref{lag.15}).  We write them in tetrad formalism \cite{Landau_2}: $\eta^{\alpha\beta}{\cal P}_\alpha{\cal P}_\beta=0$, $\eta^{\alpha\beta}{\cal P}_\alpha\pi_\beta=0$ and  
$\eta^{\alpha\beta}\pi_\alpha\pi_\beta=0$, with the Minkowski metric  $\eta^{\alpha\beta}=(-, +, +, +)$. Using rotations and Lorentz boosts, we can choose the coordinate system, where ${\cal P}_\alpha$ and $\pi_\alpha$, satisfying these constraints,  acquire the form 
\begin{eqnarray}\label{lag.20}
{\cal P}_\alpha=({\cal P}_0,  \epsilon{\cal P}_0, 0, 0), \quad  \pi_\alpha=(\pi_0, \epsilon\pi_0, 0, 0), 
\end{eqnarray}
where $\epsilon=\pm 1$ is the sign of ${\cal P}_0$.  Taking this into account, we conclude that our constraints have the following four solutions: (a) ${\cal P}_\alpha=\pi_\alpha=0$, (b) ${\cal P}_\alpha=0$, $\pi^2=0$, (c) $\pi_\alpha=\sigma{\cal P}_\alpha$, ${\cal P}^2=0$, and (d) $\pi_\alpha=0$, ${\cal P}^2=0$. Contracting these equalities with tetrad field, we conclude that they remain valid for the curve indexes as well. In accordance with these solutions, our theory consist of four sectors. Let us analyze them one by one. 

(a) ${\cal P}_\mu=\pi_\mu=0$. Together with (\ref{lag.19.1}), this implies $\dot x^0=0$, in contradiction with Eq. (\ref{best}). Therefore, in this sector our variational problem (\ref{lag.7}) has no solutions. 

(b) ${\cal P}_\mu=0$, $\pi^2=0$. Eq. (\ref{lag.19.2}) then reads $e_4 R_{\mu\nu\rho\sigma}\pi^\nu\pi^\rho\omega^\sigma=0$. Taking $e_4=0$, we arrive at $\dot x^0=0$ once again, that is our variational problem (\ref{lag.7}) has no solutions in this sector. 

(c) $\pi_\mu=\sigma{\cal P}_\mu$, ${\cal P}^2=0$ and ${\cal P}\omega=0$. The dynamical equations (\ref{lag.19.1})-(\ref{lag.19.4}) turn into 
\begin{eqnarray}\label{lag.21}
\dot x^\mu=(e_1+e_4\sigma){\cal P}^\mu, \quad 
\nabla{\cal P}_\mu=(e_1+e_4\sigma)\Theta_{\mu\nu}{\cal P}^\nu, \quad 
\nabla\omega^\mu=(e_2\sigma+e_4){\cal P}^\mu, \quad 
\left[\dot\sigma g_{\mu\nu}+\sigma(e_1+e_4\sigma)\Theta_{\mu\nu}\right]{\cal P}^\nu=0. 
\end{eqnarray}
They are written for 13 dynamical variables $x^\mu$, ${\cal P}_\mu$, $\omega^\mu$ and $\sigma$. If we assume $(e_1+e_4\sigma)\ne 0$, we have 16 differential equations for 13 variables, that can not be satisfied on a general background. 
Taking $(e_1+e_4\sigma)=0$, we arrive at $\dot x^0=0$.  Therefore, in this sector our variational problem (\ref{lag.7}) has no solutions. 

(d) $\pi_\mu=0$, ${\cal P}^2={\cal P}\omega=0$. In this sector we have $\Theta_{\mu\nu}=0$. As a consequence, Eqs. (\ref{lag.17}) are satisfied.  Dynamical equations of this sector are 
\begin{eqnarray}\label{lag.22}
\dot x^\mu=e_1{\cal P}^\mu, \qquad 
\nabla{\cal P}_\mu=0, \qquad 
\nabla\omega^\mu=e_4{\cal P}^\mu. 
\end{eqnarray}
This is the only region of constraints surface where the interacting theory possess solutions. Let us discuss the physical sector of this theory. 
Neither constraints nor equations of motion determine the variables $e_1$  and $e_4$.  Fixing the functions $e_1(\tau)$ and $e_4(\tau)$ in an arbitrary way,  the equations (\ref{lag.22})  turn into the normal system of Hamiltonian equations for determining the variables $x, {\cal P}$ and $\omega$. Solutions of the system will depend therefore on the arbitrary functions $e_1(\tau)$ and $e_4(\tau)$, so all  the basic variables in our theory have an ambiguous evolution. The variables with ambiguous
dynamics do not represent observable quantities, so we need to look  for a set of variables subject to closed system of equations, that does not involve $e_1(\tau)$ and $e_4(\tau)$.  

For this purpose, we introduce the polarization tensor 
\begin{eqnarray}\label{lag.23}
f_{\mu\nu}={\cal P}_\mu\omega_\nu-{\cal P}_\nu\omega_\mu.
\end{eqnarray}
As a consequence of Eqs. (\ref{lag.22}), it undergoes the parallel transport along the null geodesics  $x^\mu(\tau)$. In particular, it has the causal evolution. The variables $x^\mu$, ${\cal P}^\mu\equiv g^{\mu\nu}{\cal P}_\nu$, $f_{\mu\nu}$ and $e_1$ obey the following closed system of differential equations: 
\begin{eqnarray}\label{lag.24}
\dot x^\mu=e_1{\cal P}^\mu, \qquad \nabla{\cal P}^\mu=0, \qquad \nabla f_{\mu\nu}=0. 
\end{eqnarray}
They are accompanied by the algebraic relations
\begin{eqnarray}\label{lag.25}
{\cal P}^2=0, \qquad f_{\mu\nu}{\cal P}^\nu=0, \qquad \tilde f_{\mu\nu}{\cal P}^\nu=0,
\end{eqnarray}
implied by the basic constraints ${\cal P}^2={\cal P}\omega=0$.  The dual tensor in curved space is defined as follows: 
$\tilde f_{\mu\nu}=\sqrt{-\det g_{\mu\nu}}\epsilon_{\mu\nu\rho\sigma}f^{\rho\sigma}$, $\epsilon_{0123}=-1$. The remaining ambiguity due to $e_1$,
contained in these equations, is related with reparametrization invariance, and disappears when we
take our variables in the coordinate-time parametrization\footnote{In Minkowski space, we can find the causal variables within an arbitrary parametrization. They are the polarization $f_{\alpha\beta}$, momenta $p_\alpha$, and the orbital angular momentum $x_{[\alpha} p_{\beta]}$.} $x^i(t)$. 

Excluding ${\cal P}^\mu$ from the system (\ref{lag.24}) and (\ref{lag.25}), we obtain generally covariant equations for the trajectory
$\nabla\frac{\dot x^\mu}{e_1}=0$, $g_{\mu\nu}\dot x^\mu\dot x^\nu=0$.
They describe the null geodesic line taken in an arbitrary parametrization $\tau$, see Sect. 6.5.1 in \cite{AAD_Book}. 

So, we confirmed that physical sector of the Lagrangian theory (\ref{lag.7}), minimally interacting with gravity, coincides with that of Maxwell equations, taken in the geometrical optics approximation \cite{Frolov_2011, Dolan_2018}, the equations (\ref{lag.24}) and (\ref{lag.25}). Denote $E_i\equiv f_{i0}$, $B_i\equiv\frac12\epsilon_{ijk}f^{jk}$. In the Minkowski space the constraints (\ref{lag.25}) imply, that ${\bf p}$, ${\bf E}$ and ${\bf B}$ are mutually orthogonal and ${\bf B}=[\hat{\bf p}, {\bf E}]$, so the triad $(\hat{\bf p}\equiv{\bf p}/p^0, {\bf E}, {\bf B})$ is the right-handed, and moves with the speed of light in the direction ${\bf p}$. This, in essence, is our massless polarized particle. As we noticed in Sect. \ref{ss2}, it can be used to study the evolution of electric and magnetic fields of a plane monochromatic wave  along a chosen ray. 

We recall that the models of massive spinning particle \cite{AAD_Rec} predict the deviation of particle trajectory from a geodesic line due to the curvature-dependent contribution to the geodesic equation: $\nabla{\cal P}_\mu=-\frac14 R_{\mu\nu\rho\sigma}S^{\rho\sigma}\dot x^\nu$.  In the massless case, such a terms are considered as responsible for the gravitational spin-Hall effect \cite{Oancea_2020, Frolov_2020}.
In our model such a term, although it appears in the intermediate calculations (\ref{lag.19.2}), nevertheless disappeared from the final answer (\ref{lag.24}). Its appearance  in Eq. (\ref{lag.19.2}) is an inevitable consequence of the covariantization (\ref{lag.5}), but it vanishes in the region (d) of the constraints surface, where our theory is a self-consistent. In the result, the minimal interaction of the polarization $\omega$ with gravity  does not alter the  trajectory of our particle, that still remains the null geodesics, $\nabla{\cal P}_\mu=0$, ${\cal P}^2=0$.  This is also consistent with Maxwell equations. When formulating the Maxwell equations in curved space, we do not need to covariantize the derivatives in skew-symmetric product $F_{\mu\nu}=\partial_\mu A_\nu-\partial_\nu A_\mu$. As a consequence, in the geometrical optics approximation, light rays propagate along the null geodesics \cite{Misner_2011, Frolov_2011}. 

For completeness, we present the physical content of the sectors (b) and (c) of our model in Minkowski space. Here the equations (\ref{lag.19.1})-(\ref{lag.19.4}) read
\begin{eqnarray}\label{lag26}
\dot x^\alpha=e_1p^\alpha+e_4\pi^\alpha, \qquad \dot p^\alpha=0, \qquad \dot\omega^\alpha=e_2\pi^\alpha+ e_4 p^\alpha, \qquad \dot \pi^\alpha=0.  
\end{eqnarray}

In the sector (b) we have $\dot x^\alpha=e_4\pi^\alpha$, $\dot \pi^\alpha=0$, $\dot f_{\alpha\beta}=0$, $\pi^2=0$, where $f_{\alpha\beta}=\pi_{[\alpha}\omega_{\beta]}$. These equations  describe massless particle which carries the constant tensor $f_{\alpha\beta}$ without any special properties. 

In the sector (c) we have $\dot x^\alpha=(e_1+e_4\sigma)p^\alpha$, $\dot p^\alpha=0$, $\dot f_{\alpha\beta}=0$, $\dot\sigma=0$, $p^2=f_{\alpha\beta}p^\beta=\tilde f_{\alpha\beta}p^\beta =0$, where $f_{\alpha\beta}=p_{[\alpha}\omega_{\beta]}$. These equations  describe massless polarized particle with an extra scalar degree of freedom $\sigma$.

{\bf Equations of motion in the parameterization of coordinate time.} To complete the analysis of physical sector of the model (\ref{lag.24}), (\ref{lag.25}), let us  find a complete set of independent dynamical variables with unambiguous dynamics. We present all the details of this calculation, with the aim to show a peculiar property of the massless particle, that will be crucial when discussing non minimal interactions. 

We use the equation ${\cal P}^2=0$ to exclude ${\cal P}^0$, representing  it through the remaining variables $\boldsymbol{\cal P}=({\cal P}^1, {\cal P}^2, {\cal P}^3)$
\begin{eqnarray}\label{lag27}
{\cal P}^0={\bf g}\boldsymbol{\cal P}+\frac{\sqrt{\boldsymbol{\cal P}\gamma\boldsymbol{\cal P}}}{\sqrt{-g_{00}}}. 
\end{eqnarray}
In all equations below, the notation ${\cal P}^0$ means this function of $\boldsymbol{\cal P}$.
By ${\bf g}$ and $\gamma$ we denoted components of $(3+1)$\,-decomposition of the original metric: $g_i=-\frac{g_{0i}}{g_{00}}$, $\gamma_{ij}=g_{ij}-\frac{g_{0i}g_{0j}}{g_{00}}$, see Sect. 84 in \cite{Landau_2}.  Further, using the reparametrization covariance of Eqs. (\ref{lag.24}) and  (\ref{lag.25}), we take the coordinate time $t$ as the parameter, $\tau=t$. Then the equation for $x^0$ can be used to represent $e_1$ as $e_1=c/{\cal P}^0$. Note that $e_1>0$ implies ${\cal P}^0>0$. For the remaining positions we have $dx^i/dt=c{\cal P}^i/{\cal P}^0$. This prompts to work with a reducible set of variables, leading  to separation of variables in the resulting system. We replace
\begin{eqnarray}\label{lag28}
({\cal P}^1, {\cal P}^2, {\cal P}^3) \rightarrow (\tilde\omega, \hat{\cal P}^1, \hat{\cal P}^2, \hat{\cal P}^3), \qquad 
\tilde\omega=c{\cal P}^0>0, \qquad \hat{\cal P}^i=\frac{{\cal P}^i}{{\cal P}^0}.
\end{eqnarray}
As a consequence of (\ref{lag27}), the variables $\hat{\cal P}^i$ obey the identity 
\begin{eqnarray}\label{lag29}
{\bf g}\boldsymbol{\hat{\cal P}}+\frac{\sqrt{\boldsymbol{\hat{\cal P}}\gamma\boldsymbol{\hat {\cal P}}}}{\sqrt{-g_{00}}}=1,
\end{eqnarray}
so only two of them are independent, and Eq. (\ref{lag28}) is an invertible change of variables. It is convenient also to denote 
\begin{eqnarray}\label{lag29.1}
\hat{\cal P}^\mu=(1, \hat{\boldsymbol{\cal P}}). 
\end{eqnarray}
For the new variables, the equations (\ref{lag.24}) and  (\ref{lag.25}) imply
\begin{eqnarray}\label{lag.30}
\nabla_t\tilde\omega=0, \qquad \qquad \qquad \qquad \qquad \qquad 
\end{eqnarray}
\begin{eqnarray}\label{lag.31}
\frac{d{\bf x}}{dt}=c\boldsymbol{\hat{\cal P}}, \qquad \qquad \qquad \qquad \qquad 
\end{eqnarray}
\begin{eqnarray}\label{lag.32}
\nabla_t\boldsymbol{\hat{\cal P}}-c\boldsymbol{\hat{\cal P}}\Gamma^0{}_{\mu\nu}\hat{\cal P}^\mu\hat{\cal P}^\nu=0, \quad \mbox{or} \quad 
\frac{d\hat{\cal P}^i}{dt}+c(\Gamma^i{}_{\mu\nu}-\hat{\cal P}^i\Gamma^0{}_{\mu\nu})\hat{\cal P}^\mu\hat{\cal P}^\nu=0, 
\end{eqnarray}
\begin{eqnarray}\label{lag.33}
\nabla_tf_{\mu\nu}=0, \qquad \qquad \qquad \qquad \qquad \qquad 
\end{eqnarray}
\begin{eqnarray}\label{lag.34}
f_{\mu\nu}\hat{\cal P}^\nu=\tilde f_{\mu\nu}\hat{\cal P}^\nu=0. \qquad \qquad \qquad \qquad 
\end{eqnarray}
Note that $\tilde\omega$ does not enter into equations for other variables. 

As a consequence of (\ref{lag29}), velocity of the particle obeys the relation 
${\bf g}\frac{d{\bf x}}{dt}+\frac{1}{\sqrt{-g_{00}}}\sqrt{\frac{d{\bf x}}{dt}\gamma\frac{d{\bf x}}{dt}}=c$. This implies that speed of the photon, measured in the laboratory of a static observer, is equal to the speed of light \cite{Landau_2}. In space-time with static metric, $g_{0i}=0$, this relation acquires more transparent form: $g_{ij}\frac{dx^i}{dt}\frac{dx^j}{dt}=-c^2g_{00}$. 

The equations (\ref{lag.30})-(\ref{lag.33}) form a Hamiltonian system, therefore the Cauchy problem for them has unique solution. Dirac's procedure guarantees that any solution will satisfy the constraints (\ref{lag29}) and (\ref{lag.34}) at all instants, if they are satisfied at the initial instant of time. So the constraints can be taken into account by an  appropriate choice of the initial conditions. Taking this into account, we conclude that physical sector of the theory (\ref{lag.7}) consist of eight independent  Dirac observables. Two of them are contained in $f_{\mu\nu}$ and describe the polarization degrees of freedom. Five independent observables 
among ${\bf x}$ and $\hat{\boldsymbol{\cal P}}$ describe the position and the direction of motion of the photon. Besides, there is one more observable $\tilde\omega=c{\cal P}^0$. It is presented also in the theory of massless particle without polarization, see Appendix 1.  Thus, the number of observables in the theory of a massless particle is one more than the number of variables with physical interpretation. In the theory of a massless point particle, there is no room for the physical interpretation of this observable. However, such a possibility exists in the case of a massless polarized particle, associated with a plane wave. 

Indeed, note that according to the relations (\ref{lag28}) and (\ref{lag29}), in the flat space limit the variables $\tilde\omega$ and $\hat{\bf p}$ are related with the original variables as follows: $\hat{\bf p}={\bf p}/|{\bf p}|$, $\tilde\omega=c|{\bf p}|$. So the unit vector $\hat{\bf p}$ is normal to the wave front, while $\tilde\omega$ can be identified with the frequency of the wave, that represents our particle. In other words, we identify ${\cal P}^\mu$ with the four-dimensional wave vector (\ref{pol.2.1}).   We assume this interpretation of ${\cal P}^0$ in the next section. This  fixes the dimension of conjugate momenta of a massless particle: $[{\cal P}^\mu]=1/cm$.

\section{Non minimal interaction of polarized particle with spacetime curvature.}\label{ss4}

The inclusion  of an interaction, preserving the physical sector of a free theory with Dirac constraints, generally represents a non trivial task \cite{Joon-Hwi Kim_2021, Mondal_2020, Aleksandrov_2020, Gnatenko_2020, Bruno_2021, Kang_Wang_2017, Singh_2020, Biswas_2020_2, Everton_2020, Schneider_2020, Bosso_2021_1, Bosso_2021_2}.  Here we discuss the possibility to switch on a non minimal interaction of the massless  polarized particle with gravity within the Hamiltonian variational problem (\ref{lag.12}).  The interaction must be introduced so that the (deformed) constraints still admite the subsurface $\pi_\nu={\cal P}^2={\cal P}\omega=0$  as one of the solutions. It is remarkable, that this condition is fulfilled when any function $\pi_\mu\Omega^\mu(x, {\cal P}, f)$, with $\Omega$  orthogonal to ${\cal P}$
\begin{eqnarray}\label{non.0}
{\cal P}_\mu\Omega^\mu=0, 
\end{eqnarray}
is added to the Hamiltonian. Indeed, let us consider the variational problem
\begin{eqnarray}\label{non.1}
S_H=\int d\tau ~  p\dot x+\pi\dot\omega-
\left[\frac{e_1}{2}\left({\cal P}^2+2\pi_\mu\Omega^\mu(x, {\cal P},  f)\right)+\frac{e_2}{2}\pi^2+e_4{\cal P}\pi+\frac{e_3}{2\omega^2}({\cal P}\omega)^2\right]. 
\end{eqnarray}
This implies the constraints
\begin{eqnarray}\label{non.2}
{\cal P}^2+2\pi_\mu\Omega^\mu=0,  \quad {\cal P}\pi=0, \quad \pi^2=0,  \quad  {\cal P}\omega=0,
\end{eqnarray}
the equations of motion
\begin{eqnarray}\label{non.3}
\dot x^\mu=e_1{\cal P}^\mu+e_4\pi^\mu+e_1\pi_\nu\{x^\mu,\Omega^\nu\},  \qquad \qquad \cr 
\nabla{\cal P}_\mu=\Theta_{\mu\nu}(e_1{\cal P}^\nu+e_4\pi^\nu)+e_1\pi_\nu\{{\cal P}^\mu,\Omega^\nu\}, \quad \cr
\nabla\omega^\mu=e_2\pi^\mu+e_4{\cal P}^\mu+e_1\Omega^\mu+e_1\pi_\nu\{\omega^\mu,\Omega^\nu\},  \cr 
\nabla\pi_\mu=e_1\pi_\nu\{\pi_\mu,\Omega^\nu\}, \qquad \qquad  \qquad \qquad \qquad     
\end{eqnarray}
and the third-stage equations
\begin{eqnarray}\label{non.4}
e_1(\omega\Theta{\cal P}+({\cal P}\Omega))+e_1\pi_\nu\{{\cal P}\omega, \Omega^\nu\}+e_4\omega\Theta\pi=0, \qquad
e_1\pi\Theta{\cal P}+e_1\pi_\nu\{{\cal P}\pi, \Omega^\nu\}=0, \cr
-2e_4\pi\Theta{\cal P}+\pi_\nu\{\Omega^\nu, e_2\pi^2+2e_4{\cal P}\pi\}=0. \qquad \qquad  \qquad \qquad
\end{eqnarray}
In the sector (d) of the constraints surface, the equations (\ref{non.4}) are satisfied, while equations of motion for the variables $x^\mu, {\cal P}_\mu$ and $f_{\mu\nu}={\cal P}_{[\mu}\omega_{\nu]}$ are (\ref{lag.24}) and (\ref{lag.25}), except the dynamical equation for $f$, that now reads
\begin{eqnarray}\label{non.5}
\nabla f_{\mu\nu}=e_1{\cal P}_{[\mu}\Omega_{\nu]}.
\end{eqnarray}
Hence the parallel transport of $f$ is disturbed by non minimal interaction.

Let us discuss a number of specific examples. If we restrict ourselves with the linear on curvature and polarization interactions,  the terms with desired property (\ref{non.0}) are $\Omega^\mu\sim R^\mu{}_{\nu\rho\delta}{\cal P}^\nu \tilde f^{\rho\delta}$, and $\Omega^\mu\sim R^\mu{}_{\nu\rho\delta}{\cal P}^\nu f^{\rho\delta}$. Choosing the interaction with the dual tensor $\tilde f^{\mu\nu}$, the interaction term in the Hamiltonian is
\begin{eqnarray}\label{non.6}
H_{int}=\tilde\kappa e_1 \pi_\mu R^\mu{}_{\sigma\rho\delta}{\cal P}^\sigma \tilde f^{\rho\delta}, 
\end{eqnarray}
where $\tilde\kappa$ is a coupling constant. Then Eq. (\ref{non.5}) reads 
\begin{eqnarray}\label{non.7}
\nabla f_{\mu\nu}=e_1\tilde\kappa{\cal P}_{[\mu}R_{\nu]\sigma\rho\delta}{\cal P}^\sigma \tilde f^{\rho\delta}.
\end{eqnarray}
In the coordinate-time parameterization we obtain the equations (\ref{lag.30})-(\ref{lag.32}), (\ref{lag.34}) and 
\begin{eqnarray}\label{non.8}
\nabla_t f_{\mu\nu}=\tilde\kappa\tilde\omega\hat{\cal P}^\varphi g_{\varphi[\mu}R_{\nu]\sigma\rho\delta}\hat{\cal P}^\sigma \tilde f^{\rho\delta},
\end{eqnarray}
where, as it was combined above, $\tilde\omega=c{\cal P}^0$ represents the frequency of the photon, and $\hat{\cal P}^\sigma=(1, {\cal P}^i/{\cal P}^0)$.

We assumed that ${\cal P}^\mu$ of the  massless particle has the dimension of a wave vector, $[{\cal P}^\mu]=1/cm$. Then the coupling constant has the dimension 
$[\tilde\kappa]=1/cm^2$. Combining the dimensional constants at our disposal, we can write $\tilde\kappa=l_P^2\kappa$, where $\kappa$ is already dimensionless, and 
$l_P=\sqrt{\frac{\hbar G}{c^3}}$ is the Planck length. The linear interaction then will be very small, being suppressed by square of the Planck length $\sim 10^{-66}$ cm. Therefore we consider a non linear interactions, using the non linearity to adjust the dimension of the interaction term. We can try to divide Eq. (\ref{non.6}) on any one expression of the form: $\left(R_{\sigma\lambda\mu\nu}R^{\sigma\lambda\mu\nu}\right)^{\frac12}$, 
$\left(\nabla_\sigma R^{\sigma}{}_{\lambda\mu\nu}\nabla_\delta R^{\delta\lambda\mu\nu}\right)^{\frac13}$, 
$\left(\nabla_\delta R_{\sigma\lambda\mu\nu}\nabla^\delta R^{\sigma\lambda\mu\nu}\right)^{\frac13}$, $\ldots$ , all them are of dimension $1/cm^2$. The interaction, constructed with help of $\sqrt{R^2}$ term, does not vanish in the limit of plane space $M\rightarrow 0$, so we reject it. The second term is not appropriate, since it vanishes on-shell due to the Bianchi identity: $\nabla_\sigma R^{\sigma}{}_{\lambda\mu\nu}=\nabla_\mu R_{\lambda\nu}-\nabla_\nu R_{\lambda\mu}=0$. Using the third term, we have the interaction
\begin{eqnarray}\label{non.8.1}
H_{int}=\kappa e_1 \frac{\pi_\mu R^\mu{}_{\nu\rho\delta}{\cal P}^\nu \tilde f^{\rho\delta}}{(\nabla R, \nabla R)^{\frac13}}, 
\end{eqnarray}
where $\kappa$ is a dimensionless coupling constant. This implies the equation of motion
\begin{eqnarray}\label{non.9}
\frac{df_{\mu\nu}}{dt}=c\Gamma^\sigma{}_{\mu\rho}\hat{\cal P}^\rho f_{\sigma\nu}+c\Gamma^\sigma{}_{\nu\rho}\hat{\cal P}^\rho f_{\mu\sigma}+
\kappa\tilde\omega\frac{\hat{\cal P}^\varphi g_{\varphi[\mu}R_{\nu]\sigma\rho\delta}\hat{\cal P}^\sigma \tilde f^{\rho\delta}}{(\nabla R, \nabla R)^{\frac13}}.
\end{eqnarray}
The right hand side of this equation is a sum of torques due to minimal and non minimal interactions. 
This equation shows that parallel transport of  polarization tensor is disturbed by space-time curvature. Besides, contrary to the minimal interaction case, the  wave frequency $\tilde\omega$ now entered into the equation of motion for $f$. So, the rotation of polarization vector around the direction of light propagation in curved space is different for the photons of different frequencies, propagating along the same trajectory. Due to this, our equations predict an interesting effect that could be called an angular rainbow of light in a curved spacetime. It is discussed in the next section.

\section{Non minimal interaction with curvature of Schwarzschild spacetime: frequency-dependent rotation of the polarization plane.}\label{ss5}

Let us write an explicit form of Eq. (\ref{non.9}) in Schwarzschild spacetime and in the leading order approximation. 
We consider the Schwarzschild metric in the coordinates, where its spatial part acquires the conformally flat form \cite{Landau_2}
\begin{eqnarray}\label{sc.2}
g_{\mu\nu}dx^\mu dx^\nu=-\frac{\left(1-\frac{\alpha}{4|{\bf x|}}\right)^2}{\left(1+\frac{\alpha}{4|{\bf x|}}\right)^2}(dx^0)^2+\left(1+\frac{\alpha}{4|{\bf x|}}\right)^4d{\bf x}d{\bf x}.
\end{eqnarray}
We use the notation $\alpha=2MG/c^2$, $|{\bf x}|=\sqrt{x^i x^i}$ and $\hat x^i=\frac{x^i}{|{\bf x}|}$. In the $1/c^2$ approximation the metric  reads
\begin{eqnarray}\label{sc.3}
g_{\mu\nu}dx^\mu dx^\nu\approx\left(-1+\frac{\alpha}{|{\bf x|}}\right)(dx^0)^2+\left(1+\frac{\alpha}{|{\bf x|}}\right)d{\bf x}d{\bf x}, 
\end{eqnarray}
then the inverse metric is $g^{00}=-1-\frac{\alpha}{|{\bf x|}}$,  $g^{0i}=0$ and $g^{ij}=(1-\frac{\alpha}{|{\bf x|}})\delta^{ij}$. The non vanishing Christoffel symbols $\Gamma^\mu{}_{\nu\rho}=\frac12 g^{\mu\sigma}(\partial_\nu g_{\rho\sigma}+\partial_\rho g_{\nu\sigma}-\partial_\sigma g_{\nu\rho})$ in $1/c^2$ approximation are
\begin{eqnarray}\label{sc.4}
\Gamma^0{}_{0i}=\Gamma^i{}_{00}=\frac{\alpha}{2|{\bf x}|^2}\hat x^i, \qquad \Gamma^i{}_{jk}=-\frac{\alpha}{2|{\bf x}|^2}\left(\delta^{ij}\hat x^k+\delta^{ik}\hat x^j-\delta^{jk}\hat x^i\right). 
\end{eqnarray}
Then the non vanishing curvature components $R^\sigma{}_{\lambda\mu\nu}=\partial_\mu\Gamma^\sigma{}_{\lambda\nu}-\partial_\nu\Gamma^\sigma{}_{\lambda\mu}+\Gamma^\sigma{}_{\beta\mu}\Gamma^\beta{}_{\lambda\nu}-\Gamma^\sigma{}_{\beta\nu}\Gamma^\beta{}_{\lambda\mu}$ read 
\begin{eqnarray}\label{sc.5}
R^0{}_{i0p}=R^i{}_{00p}=-\frac{\alpha}{2|{\bf x}|^3}(\delta^{ip}-3\hat x^i \hat x^p), \qquad \qquad  \qquad  \qquad  \qquad  \qquad \cr R^i{}_{jkp}=
\frac{\alpha}{2|{\bf x}|^3}\left[\delta^{ik}(\delta^{jp}-3\hat x^j \hat x^p)-\delta^{ip}(\delta^{jk}-3\hat x^j \hat x^k)+
\delta^{jp}(\delta^{ik}-3\hat x^i \hat x^k)
-\delta^{jk}(\delta^{ip}-3\hat x^i \hat x^p)\right].  
\end{eqnarray}

Consider the algebraic relations (\ref{lag.25}). 
In curved space, the second equation from (\ref{lag.25}) does not contain metric, and for $f_{\mu\nu}$ determined in (\ref{lag.23}), this formally  implies the same consequences as in Minkowski space
\begin{eqnarray}\label{sc.6}
f_{\mu\nu}\hat{\cal P}^\nu=0 \quad \mbox{implies} \quad ({\bf E}, \hat{\boldsymbol{\cal P}})=0, \qquad {\bf E}=-[\hat{\boldsymbol{\cal P}}, {\bf B}]. 
\end{eqnarray}
Other equations in (\ref{lag.25}) involve metric, but owing to its diagonal form, their consequences have an almost Euclidean form 
\begin{eqnarray}\label{sc.7}
\epsilon_{\mu\nu\rho\sigma}\hat{\cal P}^\nu g^{\rho\alpha}g^{\sigma\beta}f_{\alpha\beta} \quad \mbox{implies} \quad (\hat{\boldsymbol{\cal P}}, {\bf B})=0, \qquad {\bf B}=-\frac{g_{11}}{g_{00}}[\hat{\boldsymbol{\cal P}}, {\bf B}], \cr
g_{\mu\nu}{\cal P}^\mu{\cal P}^\nu=0 \quad \mbox{implies} \quad (\hat{\boldsymbol{\cal P}}, \hat{\boldsymbol{\cal P}})=-\frac{g_{00}}{g_{11}}, \qquad \qquad \qquad \quad ~ 
\end{eqnarray}
then $({\bf B}, {\bf B})=\left(\frac{g_{11}}{g_{00}}\right)^2({\bf E}, {\bf E})$. Hence in the Schwarzschild metric, the set $({\bf E}, {\bf B}, \hat{\boldsymbol{\cal P}})$ in $1/c^2$ approximation can be considered as the right-handed triplet of mutually orthogonal vectors. If we neglect $1/c^2$ terms in these expressions, the vector $\hat{\boldsymbol{\cal P}}$ is of unit length, while ${\bf E}$ and ${\bf B}$ are of the same length. 

We now estimate the rate of variation of the polarization vector $E_i=f_{i0}$, implied by the equation (\ref{non.9}) in the Schwarzschild metric. Note that  $\Gamma$ and $R$ are of order $1/c^2$. So the leading contribution of terms with $\Gamma$ into Eq. (\ref{non.9}) is of order $1/c$, and  next-to-leading order contribution is of order $1/c^3$. The leading  contribution of terms with $R$ is of order $1/c^{\frac23}$, and  next-to-leading order contribution is of order $1/c^{\frac83}$. Computing the leading contributions into  Eq. (\ref{non.9}), we can use the approximation $g_{\mu\nu}=\eta_{\mu\nu}$ for all other quantities which appear in this equation. Direct computation gives the following result
\begin{eqnarray}\label{sc.8}
\frac{d{\bf E}}{dt}=\frac{MG}{c|{\bf x}|^2}\left[[\hat{\bf x}, {\bf B}]+(\hat{\bf x}, {\bf E})\hat{\boldsymbol{\cal P}}-
(\hat{\boldsymbol{\cal P}}, {\bf E})\hat{\bf x}\right]- \qquad \qquad \qquad \cr
\kappa\tilde\omega\left(\frac{12MG}{5c^2|{\bf x}|}\right)^{\frac13}\left[\left(1-(\hat{\bf x},\hat{\boldsymbol{\cal P}})^2\right)[\hat{\boldsymbol{\cal P}}, {\bf E}]-2(\hat{\bf x}, {\bf B})\hat{\bf x}+(\hat{\bf x},\hat{\boldsymbol{\cal P}})(\hat{\bf x}, {\bf B})\hat{\boldsymbol{\cal P}}\right]+\hat{\boldsymbol{\cal P}}\times(\ldots). 
\end{eqnarray}
This expression has the expected asymptotics: $\frac{d{\bf E}}{dt}\rightarrow 0$ as $|{\bf x}|\rightarrow\infty$ or $M\rightarrow 0$. We expand the total torque ${\bf T}$ (that is the right hand side of (\ref{sc.8}))  with respect to the basis $(\hat{\boldsymbol{\cal P}}, \hat{\bf E}, \hat{\bf B})$. Then the component ${\bf T}_{\cal P}$ is responsible for nutation  of the plane $({\bf E}, {\bf B})$, keeping it orthogonal to the direction of motion. The component ${\bf T}_E$ gives a variation rate of the length of ${\bf E}$. The component ${\bf T}_B$ gives a variation rate of  the polarization plane $({\bf E}, \hat{\boldsymbol{\cal P}})$ around the vector $\hat{\boldsymbol{\cal P}}$. Applying the projector $\hat B_i\hat B_j$ to Eq. (\ref{sc.8})), we obtain the explicit form of ${\bf T}_B$ as follows
\begin{eqnarray}\label{sc.9}
\left.\frac{d{\bf E}}{dt}\right|_B=-\kappa\tilde\omega\EuScript{C} ~ \sqrt[3]{\frac{12MG}{5c^2|{\bf x}|}} 
[\hat{\boldsymbol{\cal P}}, {\bf E}]\equiv[{\boldsymbol{\Omega}}, {\bf E}], \quad \mbox{where} \quad \EuScript{C}\equiv 1-(\hat{\bf x},\hat{\boldsymbol{\cal P}})^2-2(\hat{\bf x}, \hat{\bf B})^2. 
\end{eqnarray} 
As it should be \cite{Plebanski_1959, Ishihara_1988, Fayos_1982, Nouri-Zonoz_1999, Yihan_Chen_2011},  all $1/c$\,-terms of Eq. (\ref{sc.8}), originated from the minimal interaction, do not contribute into Eq. (\ref{sc.9}). The leading contribution into Faraday rotation in Schwarzschild field is exclusively due to the non minimal interaction (\ref{non.8.1}). 

Note that $(\hat{\bf x},\hat{\boldsymbol{\cal P}})^2=1$ and  $(\hat{\bf x}, \hat{\bf B})^2=0$ as $|{\bf x}|\rightarrow\infty$. At the point of perihelion ${\bf x}_p$ (the point of closest approach to the Schwarzschild center), we have $(\hat{\bf x},\hat{\boldsymbol{\cal P}})^2=0$ and $(\hat{\bf x}, \hat{\bf B})^2\leq 1$. So the angular function  $\EuScript{C}$ has the following properties: \par \noindent 
A) $\EuScript{C}\rightarrow 0$ as $|{\bf x}|\rightarrow\infty$; \par \noindent 
B) $\EuScript{C}({\bf x}_p)=+1$ if ${\bf B}$ is orthogonal to the plane of motion (${\bf E}$ is on the plane of motion); \par \noindent 
C) $\EuScript{C}({\bf x}_p)=-1$if ${\bf B}$ is on the plane of motion (${\bf E}$ is orthogonal to the plane of motion). 

Eq. (\ref{sc.9}) has clear interpretation for the typical scattering process, when an ingoing at $t=-\infty$ polarized particle is propagated towards the region of Schwarzschild field, and then it is observed in the asymptotically Minkowski region at $t=+\infty$. Eq. (\ref{sc.9}) states that  the axis ${\bf E}$ precess around the vector ${\boldsymbol{\Omega}}\sim\hat{\boldsymbol{\cal P}}$ with the angular velocity 
$|{\boldsymbol{\Omega}}|=\kappa\tilde\omega|\EuScript{C}|\sqrt[3]{12MG/5c^2|{\bf x}|}\sim 1/c^{\frac23}$.  An ingoing linearly polarized at $t=-\infty$ wave, entering into the region with non vanishing curvature, will experience the Faraday rotation of polarization plane, and appeared at $t=+\infty$ with the polarization plane rotated with respect to that of at $t=-\infty$. 

Consider the case of ingoing linearly polarized wave of given frequency, and  with the vector ${\bf E}$ on the plane of motion. Then $(\hat{\bf x}, \hat{\bf B})\approx 0$ during all scattering process, so the angular function $\EuScript{C}$ grows starting from $0$ up to $1$ when ${\bf x}\rightarrow {\bf x}_p$, and then decreases up to $0$ when ${\bf x}\rightarrow \infty$. This means that precession accumulates during evolution. In particular, this might be an important effect in study of photon spheres near the horizon of black holes \cite{Toshmatov_2020, Nucamendi_2020, Yunlong_Liu_2020, Sheoran_2020}. 

If the vector ${\bf E}$ of ingoing photon is orthogonal to the plane of motion, we have $-2(\hat{\bf x}, \hat{\bf B})=-2$ at the point of perihelion. Then the angular function has an opposite sign: $\EuScript{C}({\bf x}_p)=-1$, and the vector ${\bf E}$ will rotate in the opposite direction, as compared with the previous case. 

According to Eq. (\ref{sc.9}), the rotation angle linearly depends on the wave frequency $\tilde\omega$. 
Therefore the Schwarzschild spacetime play a role of dispersive media for the polarization axes of the waves with different frequency. Consider the linearly polarized light beam composed of photons with different frequencies but with the same polarization axis at $t=-\infty$. They will follow the same trajectory,  but leave the region with a different orientation of the polarization axes for the waves with different frequency, forming the angularly distributed rainbow, produced by the non minimal polarization-curvature interaction (\ref{non.8.1}).

\section{Conclusion.}\label{ss6}

We developed the  manifestly covariant formalism, describing a triad of mutually orthogonal vectors $({\bf p}, {\bf E}, {\bf B})$, moving with the speed of light in the direction of ${\bf p}$. The triad can be considered as a massless polarized particle, and used to capture some properties of light waves in curved spacetime. We presented two equivalent Lagrangian actions, (\ref{lag.5}) and (\ref{lag.7}), which can be used to describe the particle. The model allows for both minimal and non minimal interactions with an arbitrary gravitational field. The equations (\ref{lag.24}) and (\ref{lag.25}) of minimally interacting particle state that it propagates along the null geodesic, while the polarization tensor undergoes the parallel transport along the ray.  Hence they coincide with the Maxwell equations, taken in the geometrical optics approximation. Non minimal interaction with curvature is not unique, it can be constructed on the base of  any function $\Omega_\mu(x, {\cal P}, f)$ obeying the consistency condition (\ref{non.0}). This generally disturb the parallel transport of the polarization tensor: $\nabla f_{\mu\nu}=e_1{\cal P}_{[\mu}\Omega_{\nu]}$. To understand the meaning of this correction, we analyzed this equation for one specific choice of $\Omega_\nu$, in the leading order approximation in Schwarzschild spacetime, see Eq. (\ref{sc.9}). It predicts the Faraday rotation of light, linearly dependent on the frequency of the wave. The specific properties of the non minimal interaction can be enumerated as follows: \par
\noindent 1. Faraday rotation is presented even in the Schwarzschild spacetime. \par 
\noindent 2. It depends on the frequency, leading to an angular rainbow effect, that is to angular dispersion of polarization axes for electromagnetic waves with different frequencies, propagating along the same ray in a curved spacetime.  \par
\noindent 
3. Faraday rotation due to the curvature is of order $1/c^{\frac23}$, that is  more than due to $\Gamma$-terms in Kerr space. 

The presented model can be viewed as an attempt to go beyond the leading approximation of geometrical optics. Indeed, according to the modifications beyond the geometrical optics approximation \cite{Oancea_2020, Frolov_2020}, namely the curvature-dependent interactions could be responsible for the gravitational spin-Hall effect.  Concerning the gravitational Faraday effect, it requires to take into account the next-to-leading order approximation \cite{Fayos_1982, Ishihara_1988, Nouri-Zonoz_1999}, and hence involve derivatives of the connection.  If these contributions can be written in a covariant form, we expect that they also should depend on the curvature. So, our simple model suggests an alternative way for analysis of these effects by constructing non minimal curvature-dependent interactions. 

By construction, our model does not take into account the helicity of a photon, so it seems to be too simple to capture the spin-Hall effect. However, it is interesting that the Hall-type correction $R_{\mu\nu\rho\sigma}\pi^\rho\omega^\sigma\dot x^\nu$ inevitably appeared in the intermediate equation (\ref{lag.19.2}). 
According to the recent works \cite{Oancea_2020, Frolov_2020}, namely such a term could be responsible for the gravitational spin-Hall effect of light.
But in our model it vanishes in the region (d) of the constraints surface, where our theory is a self-consistent.
One possibility to improve this point is to relax the constraints system (\ref{lag.25}), assuming a non standard dispersion relation instead of ${\cal P}^2=0$, as it was suggested in \cite{Frolov_2011, Oancea_2020, Frolov_2020}. However, it is not clear for me, how to do this in a way, consistent with the requirement of coordinate independence of the speed of light in general relativity.

To construct the massless polarized particle with desired properties, we had to join together a number of unusual tricks within the framework of  Dirac formalism. They can be summarized as follows. \par 
\noindent 1. The constraints surface (\ref{lag.15}), (\ref{lag.15.1}) consist of four sectors, each with its own physical content. Only the sector (d) describes a massless polarized particle and admits a consistent coupling with gravity. \par 
\noindent 2. Conjugate momenta $\pi_\mu$ for $\omega^\mu$ do not participate in the construction of the physical sector, because of all them vanish on the  subsurface (d). \par 
\noindent 3. The model is based on the Hamiltonian, which contains a term quadratic in the Dirac constraint, or, in short, the non standard Hamiltonian. 
Let us  discuss this point in some details. It should be noted that some general statements on the structure of a constrained system, proved in \cite{Gitman_1990}, do not apply to this case. To illustrate this, let us compare two toy models, one with the standard and another with non standard  Hamiltonians. Consider the following first-order action on a phase space with conjugated pairs $x, p$ and $y, \pi$: 
\begin{eqnarray}\label{con.1}
S=\int d\tau ~ p\dot x+\pi\dot y-\left[H(x, p, \pi)+\lambda\pi\right], 
\end{eqnarray}
where H is a function that does not depend on $y$. Then the constraint $\pi=0$ is conserved in time, $\dot\pi=\{\pi, H+\lambda\pi\}=0$, so $\pi=0$ is the only (first class) constraint of the model. Hamiltonian equations read
\begin{eqnarray}\label{con.2}
\dot x=\frac{\partial H}{\partial p}, \qquad \dot p=-\frac{\partial H}{\partial x}, \cr
\dot y=\lambda+\frac{\partial H}{\partial\pi}, \qquad \dot\pi=0. 
\end{eqnarray}
Since the variation rate of $y$ is an arbitrary function $\lambda(\tau)$, dynamics of  $y$ is ambiguous, and it is an unobservable quantity. In a more traditional language, the variable $y$ is not invariant under the local symmetry $\delta y=\epsilon$, $\delta\lambda=\dot\epsilon$, presented in the action (\ref{con.1}). Hence the physical sector of the theory consist of the variables $x$ and $p$ with unambiguous dynamics. 
All this is in correspondence with general statements about the structure of a constrained system \cite{Dirac_1977, Gitman_1990, AAD_Book}. 

Next, let us consider the model with nonstandard Hamiltonian
\begin{eqnarray}\label{con.3}
S=\int d\tau ~ p\dot x+\pi\dot y-\left[H(x, p, \pi)+\lambda\pi^2\right], 
\end{eqnarray}
with the same function H. Once again, $\pi=0$ is the only first-class constraint of the model. But the Hamiltonian equations are now different
\begin{eqnarray}\label{con.4}
\dot x=\frac{\partial H}{\partial p}, \qquad \dot p=-\frac{\partial H}{\partial x}, \cr
\dot y=\frac{\partial H}{\partial\pi}, \qquad \dot\pi=0. 
\end{eqnarray}
In particular, there is no ambiguity in these equations at all!  This implies a different physical sector. Although the variable $\pi$ is still the first-class constraint, its conjugate  $y$ has a causal dynamics. Hence, the physical sector of the theory consist of the variables $x$, $p$ and $y$. This does not fit the general statements  \cite{Dirac_1977, Gitman_1990, AAD_Book}. The reason is that these statements were proved under the assumption that constraints of the complete theory and of its quadratic approximation have a similar structure,  see Sect. 2.2 in \cite{Gitman_1990}.  This assumption does not hold for a theory with non standard Hamiltonian. In view of the appearance of physically interesting models with a non-standard Hamiltonian, it becomes interesting the task to generalize the standard formalism of constrained systems to this case. We'll look at this issue in a future publication.

\begin{acknowledgments}
The work has been supported by the Brazilian foundation CNPq (Conselho Nacional de
Desenvolvimento Cient\'ifico e Tecnol\'ogico - Brasil),  and by Tomsk State University Competitiveness Improvement
Program.
\end{acknowledgments}

\section{Appendix 1.} 

{\bf Plane monochromatic wave.} Consider Maxwell equations in empty space 
\begin{eqnarray}\label{ap.1.1}
\partial_0 {\bf E}-[{\bf\nabla}, {\bf B}]=0, \qquad \partial_0 {\bf B}+[{\bf\nabla}, {\bf E}]=0,
\end{eqnarray}
\begin{eqnarray}\label{ap.1.2}
({\bf\nabla}, {\bf E})=0, \qquad ({\bf\nabla}, {\bf B})=0,
\end{eqnarray}
and look for their solutions of the form 
\begin{eqnarray}\label{ap.2}
{\bf E}(x^\alpha)={\bf e}_1\cos q_\alpha x^\alpha+{\bf e}_2\sin q_\alpha x^\alpha, 
\end{eqnarray}
\begin{eqnarray}\label{ap.2.1}
{\bf B}(x^\alpha)={\bf b}_1\cos q_\alpha x^\alpha+{\bf b}_2\sin q_\alpha x^\alpha.
\end{eqnarray}
Computing the derivative $\partial_0$ of equations (\ref{ap.1.1}), we find that  the Klein-Gordon equations $\Box{\bf E}=\Box{\bf B}=0$ are consequences of the Maxwell equations. Substitution of the anzatz (\ref{ap.2}) into the Klein-Gordon equation allow us to fix $q_0$ as follows:
\begin{eqnarray}\label{ap.3}
q_0^2-{\bf q}^2=0, \quad \mbox{then} \quad q_0=\epsilon|{\bf q}|, \quad \epsilon=sign(q_0)=\pm 1.
\end{eqnarray}
Substituting the anzatz (\ref{ap.2}) into Eqs. (\ref{ap.1.2}) we find that ${\bf E}$ and ${\bf B}$ lie on the plane orthogonal to ${\bf q}$
\begin{eqnarray}\label{ap.4}
({\bf e}_i, {\bf q})=0, \qquad ({\bf b}_i, {\bf q})=0, \quad \mbox{then} \quad ({\bf E}, {\bf q})=0, \qquad ({\bf B}, {\bf q})=0. 
\end{eqnarray}
Substituting the anzatz (\ref{ap.2}) into Eqs. (\ref{ap.1.1}) we obtain expression for ${\bf b}_i$ through ${\bf e}_i$, and ${\bf B}$ through ${\bf E}$
\begin{eqnarray}\label{ap.5}
{\bf b}_i=-\frac{1}{\epsilon|{\bf q}|}[{\bf q}, {\bf e}_i], \quad \mbox{then} \quad {\bf B}=-\frac{1}{\epsilon|{\bf q}|}[{\bf q}, {\bf E}].
\end{eqnarray}
Further, we note that the vectors ${\bf e}_i$ can be considered as mutually orthogonal. To see this, we rewrite (\ref{ap.2}) as follows:
\begin{eqnarray}\label{ap.6}
{\bf E}={\bf e}_1\cos(qx+\alpha-\alpha)+{\bf e}_2\sin(qx+\alpha-\alpha)={\bf c}_1\cos(qx+\alpha)+{\bf c}_2\sin(qx+\alpha),
\end{eqnarray}
where ${\bf c}_1\equiv {\bf e}_1\cos\alpha-{\bf e}_2\sin\alpha$, ${\bf c}_2\equiv {\bf e}_1\sin\alpha+{\bf e}_2\cos\alpha$. Choosing the angle $\alpha$ from the orthogonality condition $({\bf c}_1, {\bf c}_2)=0$, we find that either $\tan 2\alpha=-\frac{2({\bf e}_1, {\bf e}_2)}{{\bf e}_1^2-{\bf e}_2^2}$, or $\alpha=\pi/4$. The phase $\alpha$ in (\ref{ap.6}) can be removed by shifting the time origin, $q_0 x^0=q_0(x'^0-\alpha/q_0)$, and we arrive at the expression (\ref{ap.2}) with orthogonal vectors ${\bf c}_i$ in the place of ${\bf e}_i$. 

It is convenient to rewrite the phase factor $q_\alpha x^\alpha$ in (\ref{ap.2}) in terms of a redundant set of variables, which have a direct interpretation. For the wave with $\epsilon=+1$ we write
\begin{eqnarray}\label{ap.7}
q_\alpha x^\alpha=|{\bf q}|ct+{\bf q}{\bf x}=\frac{\tilde\omega}{c}(ct-\hat{\bf k}{\bf x}),  \qquad |\hat{\bf k}|=1,
\end{eqnarray}
that is we make the change of variables $(q_1, q_2, q_3)\rightarrow(\tilde\omega, \hat k_1, \hat k_2, \hat k_3)$ where 
$\tilde\omega=c|{\bf q}|$, $\hat k_i=-q_i/|{\bf q}|$. For the wave with $\epsilon=-1$ we write
\begin{eqnarray}\label{ap.8}
q_\alpha x^\alpha=-|{\bf q}|ct+{\bf q}{\bf x}=-\frac{\tilde\omega}{c}(ct-\hat{\bf k}{\bf x}), \qquad |\hat{\bf k}|=1, 
\end{eqnarray}
that is in this case we make the change of variables with another choice of $\hat k_i$: $(q_1, q_2, q_3)\rightarrow(\tilde\omega, \hat k_1, \hat k_2, \hat k_3)$ where $\tilde\omega=c|{\bf q}|$, $\hat k_i=q_i/|{\bf q}|$. 

Collecting these results, the plane wave solutions to Maxwell equations read
\begin{eqnarray}\label{ap.9}
{\bf E}(x^\alpha)={\bf e}_1\cos \frac{\epsilon\tilde\omega}{c}(ct-\hat{\bf k}{\bf x}) +{\bf e}_2\sin \frac{\epsilon\tilde\omega}{c}(ct-\hat{\bf k}{\bf x})\equiv  \qquad \qquad \qquad \cr 
{\bf e}_1\cos \frac{-\tilde\omega}{c}(ct-\hat{\bf k}{\bf x}) -\epsilon{\bf e}_2\sin \frac{-\tilde\omega}{c}(ct-\hat{\bf k}{\bf x})\equiv
{\bf e}_1\cos \eta_{\alpha\beta}k^\alpha x^\beta -\epsilon{\bf e}_2\sin \eta_{\alpha\beta}k^\alpha x^\beta,  \cr
{\bf B}(x^\alpha)=[\hat{\bf k}, {\bf E}], \qquad \qquad \qquad \qquad \qquad \qquad \qquad 
\end{eqnarray}
where we introduced the Lorentz-invariant expression for the phase in terms of four dimensional wave vector 
\begin{eqnarray}\label{ap.9.1}
k^\alpha=\left(\frac{\tilde\omega}{c}, \frac{\tilde\omega}{c}\hat{\bf k}\right), \qquad k^2=0. 
\end{eqnarray} 
The unit vector $\hat{\bf k}$ points the direction orthogonal to a wave front, $\tilde\omega>0$ is  frequency of the wave,  $T=2\pi/\tilde\omega$ gives period, and $\lambda=cT$ is the wavelength. The set $(\hat{\bf k}, {\bf E}, {\bf B})$  consists of mutually orthogonal vectors which form the right-handed triad, and $|{\bf E}|=|{\bf B}|$. The plane of $\hat{\bf k}$ and ${\bf E}$ is called the plane of polarization. 

If we look at the plane of ${\bf E}$ and ${\bf B}$ from the end of the vector $\hat{\bf k}$, the vectors ${\bf E}(t)$ and ${\bf B}(t)$ rotate clockwise when $\epsilon=+1$, and counterclockwise when $\epsilon=-1$. To see this, consider the plane wave in special coordinates constructed as follows. As $(\hat{\bf k}, {\bf e}_1, {\bf e}_2)$ are orthogonal constant vectors, take the coordinate system with $x$\,-axis along the vector $\hat{\bf k}$ and with $z$\,-axis along ${\bf e}_1$. In these coordinates we can write 
\begin{eqnarray}\label{ap.10}
\hat{\bf k}=\left(
\begin{array}{c}
1 \\
0 \\
0
\end{array}
\right). \qquad  \bf {e}_2=\left(
\begin{array}{c}
0 \\
\epsilon_2 e_{2y} \\
0
\end{array}
\right), \qquad  {\bf e}_1=\left(
\begin{array}{c}
0 \\
0 \\
e_{1z}
\end{array}
\right), 
\end{eqnarray}
where $e_{1z}$ and $e_{2y}$ are positive numbers and $\epsilon_2=\pm 1$. The electric field in these coordinates acquires the form 
\begin{eqnarray}\label{ap.11}
E_x=0, \qquad E_y=\epsilon_2 e_{2y}\sin\frac{\epsilon\tilde\omega}{c}(ct-x), \qquad E_z=e_{1z}\cos\frac{\epsilon\tilde\omega}{c}(ct-x), \quad \mbox{then} \quad 
\frac{E_z^2}{e_{1z}^2}+\frac{E_y^2}{e_{2y}^2}=1. 
\end{eqnarray}
The wave with this ${\bf E}$ we call the $(\epsilon, \epsilon_2)$\,-wave. Then $(+, +)$ and $(-, -)$ waves are clockwise and are indistinguishable. Similarly,
$(-,+)$ and $(+,-)$ waves are counterclockwise and are indistinguishable.  So, it is sufficient to consider only $(+, +)$ and $(-, +)$ waves, that is those with $\epsilon_2=+1$. Then the sign of $\epsilon$ determines their helicity. Besides, when $\epsilon_2=+1$, the set $(\hat{\bf k}, {\bf e}_2, {\bf e}_1)$  is the right-handed triad. 

Consider the plane wave with non vanishing ${\bf e_1}$, ${\bf e_2}$ at a fixed point ${\bf x}$. According to Eqs. (\ref{ap.11}) and (\ref{ap.9}), the ends of ${\bf E}(t)$ and ${\bf B}(t)$ lie on ellipses with semi-axes $|{\bf e}_1|$ and $|{\bf e}_2|$. They rotate in their own plane, and the angle of rotation is $\beta(t)=\arctan(E_y/E_z)$. Then the angular velocity is $d\beta/dt=\epsilon\tilde\omega\left(\frac{|{\bf e}_1|}{|{\bf e}_2|}\cos^2\frac{\epsilon\tilde\omega}{c}(ct-\hat{\bf k}{\bf x})+
\frac{|{\bf e}_2|}{|{\bf e}_1|}\sin^2 \frac{\epsilon\tilde\omega}{c}(ct-\hat{\bf k}{\bf x})\right)^{-1}$. It is a constant for circular polarization, when $|{\bf e}_1|=|{\bf e}_2|$. One revolution occurs in the period $T=2\pi/\tilde\omega$. Let us now fix $t$ and consider the straight line through ${\bf x}$ in the direction of $\hat{\bf k}$, that is we consider the instantaneous configuration of the wave along the straight line ${\bf x}+\hat{\bf k}s$, $s\in{\mathbb R}$. Then the ends of the vectors ${\bf E}$ and ${\bf B}$ lie on the surface of the elliptical cylinder, and make one revolution after the increment  $\triangle s=cT$. When one of the vectors ${\bf e}_i$ vanishes, ${\bf E}$ oscillates along the another vector, and we have the linearly-polarized wave. The linearly-polarized waves with $\epsilon=\pm 1$ are indistinguishable.

{\bf Variational problem for the massless point particle.} To describe the massless point particle, we need a variational problem producing the following set  of trajectories: 
\begin{eqnarray}\label{apv.1}
{\bf x}(t)={\bf x}_0+c\hat{\bf v}t, \qquad x_0^i, \hat v^i\in {\mathbb R}, \quad \hat{\bf v}^2=1. 
\end{eqnarray}
It turns out to be sufficient to work with the action
\begin{eqnarray}\label{apv.2}
S=\int d\tau \frac{1}{2e}\eta_{\alpha\beta}\dot x^\alpha\dot x^\beta,
\end{eqnarray}
with positively defined auxiliary variable, $e(\tau)>0$, and within the class of increasing parameterizations, $dx^0/d\tau>0$. This implies, that conjugated momentum for $x^0$ with the upper index: $p^0=\eta^{0\alpha}\frac{\partial L}{\partial\dot x^\alpha}=-\frac{\partial L}{\partial\dot x^0}$ is a positively defined function. 
To confirm these claims, consider the action
\begin{eqnarray}\label{apv.3}
S=\int dt \frac{1}{2e}(\dot{\bf x}^2-c^2),
\end{eqnarray}
with positively defined dynamical variable $e(t)$. This implies the equations 
\begin{eqnarray}\label{apv.4}
\frac{d}{dt}\left(\frac{1}{e}\frac{d{\bf x}}{dt}\right)=0, \qquad \frac{d{\bf x}}{dt}=c^2.
\end{eqnarray}
As $e(t)>0$, these equations imply  $\frac{1}{e}\frac{d{\bf x}}{dt}={\bf b}=const$, $e=c/|{\bf b}|$. Substituting this $e$ into the previous equation and integrating it, we arrive at the desired answer (\ref{apv.1}) with $\hat{\bf v}={\bf b}/|{\bf b}|$. Changing the variable of integration in (\ref{apv.3}) from $t$ to $t(\tau)$, where $t(\tau)$ is an increasing function: $dt/d\tau>0$, we obtain the reparametrization invariant action (\ref{apv.2}) with $e(\tau)\equiv\frac{dt}{d\tau}e(t(\tau))>0$. The action also implies the equations of motion with the set of solutions being (\ref{apv.1}). Besides, the action (\ref{apv.2}) immediately implies $p^0=\dot x^0/e>0$ for all inertial observers.

The theory (\ref{apv.2}) has the following peculiar property: in the Hamiltonian formulation of the model, one of the Dirac observables has no of physical interpretation. Indeed, in the  Hamiltonian formulation we have
\begin{eqnarray}\label{apv.5}
\dot x^\alpha=ep^\alpha, \qquad \dot p^\alpha=0, \qquad p^2=0.
\end{eqnarray}
Excluding $p^0$ with help of the algebraic equation, and taking $t$ as a parameter, $\tau=t$, we obtain six variables with definite dynamics: $\frac{d{\bf x}}{dt}=c\frac{{\bf p}}{|{\bf p}|}$, $\frac{d{\bf p}}{dt}=0$. This prompts to work with the reducible set of variables: $(p^1, p^2, p^3)\rightarrow (p^0, \hat p^1, \hat p^2, \hat p^3)$, where $p^0=|{\bf p}|$, $\hat p^i=p^i/|{\bf p}|$. They obey the equations $\frac{d{\bf x}}{dt}=c\hat{\bf p}$, $\frac{d\hat{\bf p}}{dt}=0$, $|\hat{\bf p}|=1$, $\frac{dp^0}{dt}=0$. The observables ${\bf x}(t)$ and $\hat{\bf p}(t)$ describe the position and the direction of motion of the light particle. The observable $p^0(\tau)$ has no an interpretation in the theory of a massless point particle. The same result follows from the Hamiltonian analysis of the action (\ref{apv.3}).

\section{Appendix 2.}

Here we construct Hamiltonian formulation for the theory (\ref{lag.5}) with three auxiliary variables
\begin{eqnarray}\label{app.1}
S=\int d\tau\frac{1}{4e_1}\left[\dot x^2+\frac{1-e_3}{1+e_3}\frac{(\dot x\omega)^2}{\omega^2}+e_2(\nabla\omega)^2-\sqrt{\left(\dot xN\dot x+e_2\nabla\omega
N\nabla\omega\right)^2-4e_2(\dot xN\nabla\omega)^2}\right], 
\end{eqnarray}
and show that this implies all the desired constraints (\ref{pol.5}) which, together with dynamical equations, arise as the conditions of extreme of this variational problem. Hence the theory describes a photon in an arbitrary gravitational background.

Conjugate momenta for $x^\mu$, $\omega^\mu$ and $e_i$ are
denoted as $p_\mu$, $\pi_\mu$ and $p_{ei}$.  Since
$p_{ei}=\frac{\partial L}{\partial\dot e_i}=0$, the momenta $p_{ei}$ represent the trivial primary constraints,
$p_{ei}=0$. Expressions for the remaining momenta, $p_\mu=\frac{\partial L}{\partial\dot x^\mu}$ and
$\pi_\mu=\frac{\partial L}{\partial\dot\omega^\mu}$, can be written in the form
\begin{eqnarray}\label{app.2}
{\cal P}_\mu=\frac{1}{2e_1}(\dot x_\mu-K_\mu+\frac{1-e_3}{1+e_3}\frac{(\dot x\omega)}{\omega^2}\omega_\mu), \qquad \qquad \qquad \cr K_\mu\equiv T^{-\frac12}\left[\left(\dot xN\dot x+e_2\nabla\omega
N\nabla\omega\right)(N\dot x)_\mu-2e_2(\dot xN\nabla\omega)(N\nabla\omega)_\mu\right],
\end{eqnarray}
\begin{eqnarray}\label{app.3}
\pi^\mu=\frac{e_2}{2e_1}(\nabla\omega^\mu-R^\mu), \qquad \qquad \qquad \cr R^\mu\equiv T^{-\frac12}\left[\left(\dot xN\dot x+e_2\nabla\omega
N\nabla\omega\right)(N\nabla\omega)^\mu-2(\dot xN\nabla\omega)(N\dot x)^\mu\right],
\end{eqnarray}
where it was denoted $T=\left(\dot xN\dot x+e_2\nabla\omega N\nabla\omega\right)^2-4e_2(\dot xN\nabla\omega)^2$. 
In these expressions also appeared the canonical momentum 
\begin{eqnarray}\label{app.4}
{\cal P}_\mu\equiv p_\mu-\pi_\rho\Gamma^\rho{}_{\mu\nu}\omega^\nu. 
\end{eqnarray}
Contrary to $p_\mu$, the canonical momentum is a four vector, so we expect that Hamiltonian and equations of motion will be written in terms of the covariant quantity ${\cal P}_\mu$. 
The functions $K^\mu$ and $R^\mu$ obey the
following identities
\begin{eqnarray}\label{app.5}
K^2=\dot xN\dot x, \quad R^2=\nabla\omega N\nabla\omega, \quad KR=-\dot x N\nabla\omega, \qquad \cr \dot xR+\nabla\omega K=0,  \qquad \dot
xK+e_2\nabla\omega R=T^{\frac12} \qquad  \omega K=\omega R=0, \cr 
\qquad \omega\dot x=\frac{2e_1}{n+1}{\cal P}\omega, \qquad \omega\nabla\omega=\frac{2e_1}{e_2}\omega\pi. \qquad \quad 
\end{eqnarray}
Computing the expression ${\cal P}N\pi$ with use of (\ref{app.2}), (\ref{app.3}) and (\ref{app.5}), we arrive at one more primary constraint
\begin{eqnarray}\label{app.6}
{\cal P}N\pi=0. 
\end{eqnarray}
Hamiltonian is obtained excluding velocities from the expression
\begin{eqnarray}\label{app.7}
H=p\dot x+\pi\dot\omega-L+\lambda({\cal P}N\pi)+\lambda_ip_{ei},
\end{eqnarray}
where $\lambda$ and $\lambda_i$ are the Lagrangian multipliers for the primary constraints.  To obtain the manifest form of $H$, we
note the equalities 
\begin{eqnarray}\label{app.8}
p\dot x+\pi\dot\omega\equiv{\cal P}\dot x+\pi \nabla\omega= 2L,  \qquad 
{\cal P}^2+\frac{1}{e_2}\left[\pi^2+\frac{(\omega\pi)^2}{\omega^2}\right]+e_3\frac{({\cal P}\omega)^2}{\omega^2}=\frac{2}{e}L,
\end{eqnarray}
that can be verified with use of  (\ref{app.2})-(\ref{app.5}). Using them in (\ref{app.7}), we obtain the Hamiltonian. The corresponding Hamiltonian action reads
\begin{eqnarray}\label{app.9}
S_H=\int d\tau ~ p\dot x+\pi\dot\omega-H=\int d\tau ~  p\dot x+\pi\dot\omega-\left[\frac{e_1}{2}{\cal P}^2+\frac{e_1}{2e_2}\left(\pi^2+\frac{(\omega\pi)^2}{\omega^2}\right)+\frac{e_1e_3}{2\omega^2}({\cal P}\omega)^2+\lambda({\cal P}N\pi)+\lambda_ip_{ei}\right]. 
\end{eqnarray}
It is accompanied by the brackets (\ref{lag.13}) and (\ref{lag.14}).

Preservation in time of primary constraints, $\dot p_{ei}=\{p_{ei}, H\}=0$, gives the algebraic equations of second stage of the Dirac procedure: ${\cal P}^2=0$, $\pi^2+\frac{(\omega\pi)^2}{\omega^2}=0$, and $\frac{({\cal P}\omega)^2}{\omega^2}=0$. They imply that all solutions of the variational problem (if any) lie on the constraints surface 
\begin{eqnarray}\label{app.12}
{\cal P}^2=0,  \quad {\cal P}\pi=0, \quad \pi^2+\frac{(\omega\pi)^2}{\omega^2}=0,  \quad {\cal P}\omega=0. 
\end{eqnarray}
The term  $\frac{e_1e_3}{2\omega^2}({\cal P}\omega)^2$ is a square of the constraint $({\cal P}\omega)=0$, and can be omitted from the Hamiltonian, that reads
\begin{eqnarray}\label{app.13}
H=\frac{e_1}{2}\left[{\cal P}^2+\frac{1}{e_2}\left(\pi^2+\frac{(\omega\pi)^2}{\omega^2}\right)\right]+\lambda({\cal P}N\pi)+\lambda_ip_{ei}. 
\end{eqnarray}

For the latter use, let us obtain the algebraic equations of third stage of the Dirac procedure. The constraint 
$\pi^2+(\omega\pi)^2/\omega^2=0$ is of first class, so it automatically preserved in time. For the remaining constraints we have
\begin{eqnarray}\label{app.14}
\{{\cal P}\omega, H\}=0 ~\Rightarrow ~ e_1\omega\Theta{\cal P}+\lambda\omega\Theta\pi=0, \cr
\{{\cal P}\pi, H\}=0 ~\Rightarrow ~ e_1\pi\Theta{\cal P}+\lambda\frac{\omega\pi}{\omega^2}\omega\Theta\pi=0, \cr
\{{\cal P}^2, H\}=0 ~\Rightarrow ~ \lambda(\pi\Theta{\cal P}-\omega\Theta{\cal P})=0.
\end{eqnarray}
The last equation is a consequence of the other two, so it can be omitted from the consideration. The remaining two equations can be presented in an equivalent form as follows: 
\begin{eqnarray}\label{app.15}
e_1\omega\Theta{\cal P}+\lambda\omega\Theta\pi=0, \qquad e_1{\cal P}\Theta N\pi=0,
\end{eqnarray}
that is we have an equation for determining the Lagrangian multiplier $\lambda$, and a new constraint. 
These equations are absent in Minkowski space, and their appearance in interacting theory would mean its inconsistency. 
As we show below, the constraints surface (\ref{app.12}) consist of three regions. Only in one of them our theory admits a self-consistent interaction with an arbitrary gravitational field. In this region, the equations (\ref{app.15}) will be  automatically satisfied. 

Computing the Hamiltonian equations $dq/d\tau=\{ q, H\}$,we obtain
\begin{eqnarray}\label{app.17}
\dot x^\mu=e_1{\cal P}^\mu+\lambda N\pi^\mu, \qquad \nabla{\cal P}_\mu=\Theta_{\mu\nu}(e_1{\cal P}^\nu+\lambda N\pi^\nu)\equiv\Theta_{\mu\nu}\dot x^\nu, \cr
\nabla\omega^\mu=\frac{e_1}{e_2}\left(\pi^\mu+\frac{\omega\pi}{\omega^2}\omega^\mu\right)+\lambda{\cal P}^\mu, \qquad \nabla\pi_\mu=-\frac{\omega\pi}{\omega^2}\left[\frac{e_1}{e_2}N\pi_\mu-\lambda{\cal P}_\mu\right]. 
\end{eqnarray}

To analyze the resulting theory, it is convenient to rewrite all the equations in terms of variables $x^\mu, {\cal P}_\mu, \omega^\mu, \Pi_\mu$, where
\begin{eqnarray}\label{app.18} 
\Pi_\mu=\pi_\mu-(1+\sqrt 2)\frac{\omega\pi}{\omega^2}\omega_\mu, \quad \mbox{then} \quad \omega\pi=-\frac{1}{\sqrt 2}\omega\Pi, \qquad \pi_\mu=\Pi_\mu-\frac{1+\sqrt 2}{\sqrt 2}\frac{\omega\Pi}{\omega^2}\omega_\mu. 
\end{eqnarray}
Then the constraints (\ref{app.12}) acquire a more transparent form 
\begin{eqnarray}
{\cal P}^2=0,  \qquad {\cal P}\Pi=0, \qquad \Pi^2=0,  \label{app.19.1} \\ 
{\cal P}\omega=0, \qquad \qquad \qquad  \label{app.19.2}
\end{eqnarray}
while the dynamical equations (\ref{app.17}) turn into 
\begin{eqnarray}
\dot x^\mu=e_1{\cal P}^\mu+\lambda N\Pi^\mu, \label{app.19}  \qquad \qquad \qquad \qquad \qquad \quad \\  
\nabla{\cal P}_\mu=\Theta_{\mu\nu}(e_1{\cal P}^\nu+\lambda N\Pi^\nu), \label{app.20} \qquad \qquad \qquad \qquad \\ 
\nabla\omega^\mu=\frac{e_1}{e_2}\left(\Pi^\mu-(1+\sqrt 2)\frac{\omega\Pi}{\omega^2}\omega^\mu\right)+\lambda{\cal P}^\mu,  \label{app.21} \qquad \\ 
\nabla\Pi_\mu=\frac{\omega\Pi}{\sqrt 2\omega^2}\left[\frac{e_1}{e_2}(2+\sqrt 2)
\Pi_\mu+(1+\sqrt 2)\lambda{\cal P}_\mu\right], \label{app.22}
\end{eqnarray}
where now $\Theta_{\mu\nu}\equiv R_{\mu\nu\rho\delta}\Pi^\rho\omega^\delta$. Let us analyse the constraints (\ref{app.19.1}).  We write them in tetrad formalism: $\eta^{\alpha\beta}{\cal P}_\alpha{\cal P}_\beta=0$, $\eta^{\alpha\beta}{\cal P}_\alpha\Pi_\beta=0$ and  
$\eta^{\alpha\beta}\Pi_\alpha\Pi_\beta=0$, with the Minkowski metric  $\eta^{\alpha\beta}=(-, +, +, +)$. Using rotations and Lorentz boosts, we can choose the coordinate system, where ${\cal P}_\alpha$ and $\Pi_\alpha$, satisfying these constraints,  acquire the form 
\begin{eqnarray}\label{app.24}
{\cal P}_\alpha=({\cal P}_0,  \epsilon{\cal P}_0, 0, 0), \quad  \Pi_\alpha=(\Pi_0, \epsilon\Pi_0, 0, 0), 
\end{eqnarray}
where $\epsilon=\pm 1$ is the sign of ${\cal P}_0$.  Taking this into account, we conclude that our constraints have the following four solutions: (a) ${\cal P}_\alpha=\Pi_\alpha=0$, (b) ${\cal P}_\alpha=0$, $\Pi^2=0$, (c) $\Pi_\alpha=\sigma{\cal P}_\alpha$, ${\cal P}^2=0$, and (d) $\Pi_\alpha=0$, ${\cal P}^2=0$. Contracting these equalities with tetrad field, we conclude that they remain valid for the curve indexes as well. In accordance with these solutions, our theory consist of four sectors. Let us analyze them one by one. 

(a) ${\cal P}_\mu=\Pi_\mu=0$. Together with (\ref{app.19}), this implies $\dot x^0=0$, in contradiction with Eq. (\ref{best}). Therefore, in this sector our variational problem (\ref{app.1}) has no solutions. 

(b) ${\cal P}_\mu=0$, $\Pi^2=0$. Eq. (\ref{app.20}) then reads $\lambda R_{\mu\nu\rho\sigma}(N\Pi)^\nu\Pi^\rho\omega^\sigma=0$. Taking either $\lambda=0$ or $(N\Pi)^\nu=0$, we arrive at $\dot x^0=0$ once again, that is our variational problem (\ref{app.1}) has no solutions in this sector. 

(c) $\Pi_\mu=\sigma{\cal P}_\mu$, ${\cal P}^2=0$ and ${\cal P}\omega=0$. This implies $\omega\Pi=\sigma\omega{\cal P}=0$, so the dynamical equations (\ref{app.19})-(\ref{app.22}) turn into 
\begin{eqnarray}\label{app.25}
\dot x^\mu=(e_1+\lambda\sigma){\cal P}^\mu, \quad 
\nabla{\cal P}_\mu=(e_1+\lambda\sigma)\Theta_{\mu\nu}{\cal P}^\nu, \quad 
\nabla\omega^\mu=(\frac{e_1}{e_2}\sigma+\lambda){\cal P}^\mu, \quad 
\left[\dot\sigma g_{\mu\nu}+\sigma(e_1+\lambda\sigma)\Theta_{\mu\nu}\right]{\cal P}^\nu=0. 
\end{eqnarray}
They are written for 13 dynamical variables $x^\mu$, ${\cal P}_\mu$, $\omega^\mu$ and $\sigma$. If we assume $(e_1+\lambda\sigma)\ne 0$, we have 16 differential equations for 13 variables, that can not be satisfied on a general background. 
Taking $(e_1+\lambda\sigma)=0$, we arrive at $\dot x^0=0$.  Therefore, in this sector our variational problem (\ref{app.1}) has no solutions. 

(d) $\Pi_\mu=0$, ${\cal P}^2={\cal P}\omega=0$. In this sector we have $\Theta_{\mu\nu}=0$. As a consequence, the constraints ${\cal P}^2=0$ and ${\cal P}\omega=0$ are of first class. Besides,  Eqs. (\ref{app.15}) are satisfied. We note also that Eq. (\ref{app.18}) implies $\pi_\mu=0$. Dynamical equations of this sector are 
\begin{eqnarray}\label{app.26}
\dot x^\mu=e_1{\cal P}^\mu, \qquad 
\nabla{\cal P}_\mu=0, \qquad 
\nabla\omega^\mu=\lambda{\cal P}^\mu. 
\end{eqnarray}
They coincide with the equations (\ref{lag.22}), with the auxiliary variable $e_4$ renamed as $\lambda$. Continuing the analysis, as it was done at the end of  Sect. \ref{lagrangian}, we arrive at the equations (\ref{lag.24}) and (\ref{lag.25}).

\end{document}